\documentclass[lettersize,journal]{IEEEtran}
\usepackage{amsmath,amsfonts}
\usepackage{array}
\usepackage[caption=false,font=normalsize,labelfont=sf,textfont=sf]{subfig}
\usepackage{textcomp}
\usepackage{stfloats}
\usepackage{url}
\usepackage{verbatim}
\usepackage{graphicx}
\usepackage{cite}
\usepackage{cite}
\usepackage{amsthm}

\usepackage{amsmath}  
\usepackage{cases} 
\usepackage{bm}
\usepackage{multirow} 
\usepackage{multicol}
\usepackage{diagbox}
\usepackage{mathtools} 
\usepackage{subfig}
\usepackage{float}
\usepackage{cases}
\usepackage{xcolor}  
\usepackage{booktabs}
\usepackage{ulem}
\usepackage{varwidth}
\usepackage{hyperref}
\hypersetup{hypertex=true,citebordercolor=blue,linkcolor=black}
\usepackage{enumitem}
\usepackage{balance}

\newtheorem{myDef}{Definition}

\usepackage{algorithm}
\usepackage{algorithmic}

\usepackage{newfloat}
\usepackage{listings}

\begin{document}

\title{CogGen: Cognitive-Load-Inspired Fully Unsupervised Deep Generative Modeling for Compressively Sampled MRI Reconstruction}

\author{
	Qingyong Zhu, Yumin Tan, Xiang Gu and Dong Liang
	\thanks{This work was supported in part by the National Natural Science Foundation of China under Grants 62125111 and 62331028; in part by the Guangdong Basic and Applied Basic Research Foundation under Grant 2023A1515110476; and in part by the Guangdong Provincial Key Laboratory of Multimodality Non-Invasive Brain-Computer Interfaces under Grant 2024B1212010010. (Corresponding author: Dong Liang.)}
	\thanks{Qingyong Zhu is with the Medical AI Research Centre, Shenzhen Institutes of Advanced Technology, Chinese Academy of Sciences, Shenzhen 518055, Guangdong, China (qy.zhu@siat.ac.cn).}
	\thanks{Yumin Tan is with the School of Biomedical Engineering, Southern Medical University, Guangzhou 510515, Guangdong, China.}
	\thanks{Xiang Gu is with the School of Mathematics and Statistics, Xi'an Jiaotong University, Xi'an 710049, Shaanxi, China.}
	\thanks{Dong Liang is with the Paul C. Lauterbur Research Centre for Biomedical Imaging, Shenzhen Institutes of Advanced Technology, Chinese Academy of Sciences, Shenzhen 518055, Guangdong, China (dong.liang@siat.ac.cn). }
}


\markboth{Journal of \LaTeX\ Class Files,~Vol.~14, No.~8, August~2021}%
{Shell \MakeLowercase{\textit{et al.}}: A Sample Article Using IEEEtran.cls for IEEE Journals}


\maketitle

\begin{abstract}
	
\noindent Fully unsupervised deep generative modeling (FU-DGM) offers significant potential for compressively sampled magnetic resonance imaging (CS-MRI) reconstruction. Representative FU-DGM formulations, such as deep image prior (DIP) and implicit neural representation (INR), employ architectural bias to induce a low-dimensional manifold in the image space that aligns with the forward observation. However, as the underlying inverse system is highly ill-posed, prolonged iterative fitting in FU-DGM typically leads to poor efficiency and noise amplification. In this paper, guided by the cognitive principle of easy-to-hard learning, we propose CogGen, an FU-DGM framework that reformulates CS-MRI reconstruction as a staged inversion problem. Specifically, CogGen implements an self-paced curriculum learning (SPCL)-driven progressive scheduling strategy through an MRI-aware dual-threshold weighting criterion, which adaptively regulates k-space measurement participation. The data-consistency residual thresholding evaluates the fitting reliability of the current generator, while the k-space radius thresholding controls stage-wise measurement exposure, thereby avoiding uniform fitting throughout optimization. Theoretically, our analysis shows that, when early stages favor easy-to-fit measurements, CogGen yields a reduced local sufficient-iteration bound and a smaller cumulative noise-amplification bound, explaining the improved convergence behavior and reconstruction fidelity of CogGen within a finite iteration budget. Numerical experiments demonstrate that both CogGen instantiations, CogGen-DIP and CogGen-INR, achieve superior performance over prevailing CS-MRI reconstruction techniques, including unsupervised and supervised pipelines.

\end{abstract}


\section{Introduction}
\indent \IEEEPARstart{M}agnetic resonance imaging (MRI) is a leading medical-imaging modality that noninvasively visualizes internal tissues, providing critical information for assessing not only anatomical structures but also pathological conditions~\cite{hussain2022modern, de2024role}. However, protracted scan times due to hardware limitations impede scanner throughput and patient comfort, thereby hindering the deployment of advanced clinical applications such as multi-contrast~\cite{kits20242} and real-time imaging~\cite{guo2022emerging}. Consequently, compressively sampled MRI (CS-MRI)~\cite{sandino2020compressed,curatolo2022recent,tavakkoli2025review}, in which k-space data are deliberately acquired below the Nyquist rate, has long been a central topic of investigation; yet achieving accurate and reliable image reconstruction from such incomplete measurements remains a substantial challenge.\\
\indent In recent years, deep generative modeling (DGM)~\cite{ruthotto2021introduction,suzuki2022survey} has attracted increasing attention for CS-MRI reconstruction, frequently demonstrating notable advantages over conventional baselines based on variational or iterative regularization~\cite{eksioglu2016decoupled,2016Infimal,cohen2021regularization,zhu2023characteristic} through a generative parameterization that implicitly encodes a low-dimensional image manifold whose elements are consistent with the acquired k-space measurements. Existing DGM-based methods can be broadly categorized into two families. The first family~\cite{8233175,8327637,li2021high} employs supervised DGM (S-DGM), wherein generators trained on extensive corpora pairing measurements with the corresponding ground-truths (GTs) learn a conditional generative mapping. Nevertheless, the dependence on task-matched, paired data imposes real-world constraints, particularly in specialized or resource-limited settings where GT labels are scarce, costly, or unattainable. The second family~\cite{9173689,yoo2021time,darestani2021accelerated,shen2022nerp,10374196}, namely unsupervised DGM (U-DGM), dispenses with paired datasets. In particular, fully unsupervised DGM (FU-DGM) represents a notable subcategory that requires no external training data and reconstructs each target image by exploiting the architectural bias of an untrained generator, thereby implicitly encoding image regularities such as spatial continuity, local correlations, and multiscale structures. Although classical FU-DGM formulations such as deep image prior (DIP)~\cite{ref26} and implicit neural representation (INR)~\cite{molaei2023implicit} yield visually plausible reconstructions, they often suffer from suboptimal accuracy, especially in fine details, due to the amplification of noise corruption in ill-posed systems~\cite{benfenati2025early}. Additionally, the prolonged iterative procedures required by such systems lead to low inversion efficiency. Existing improvements typically stabilize FU-DGM reconstruction by refining the generator architecture~\cite{heckel2018deep,xu2025self}, introducing explicit regularization~\cite{ref43,ref45} or applying early stopping~\cite{wang2021early}. Although these strategies can alleviate noise amplification or overfitting to some extent, they still provide limited control over the accuracy-efficiency tradeoff in highly ill-posed CS-MRI reconstruction. \\
\indent To address the above issues, inspired by cognitive-load theory~\cite{sweller2023development,chen2023cognitive}, which formalizes the easy-to-hard learning principle in human cognition, we introduce a novel FU-DGM framework termed CogGen. CogGen reformulates CS-MRI reconstruction as a staged inversion problem, in which the cognitive load imposed by heterogeneous k-space measurements, including sample-specific difficulty and noise-induced interference, is progressively scheduled to enhance the efficiency and robustness of scan-specific parameterization. Operationally, progressive scheduling is instantiated through a self-paced curriculum learning (SPCL) scheme~\cite{kim2018screenernet,zhang2019leveraging,choi2025brain} featuring an MRI-aware dual-threshold weighting criterion. Specifically, data-consistency residual thresholding evaluates the fitting reliability of the current generator, whereas k-space radius thresholding governs stage-wise measurement exposure. Explicitly regulating what to fit and when during reconstruction avoids uniform fitting throughout optimization, allowing early iterations to focus on structurally informative, low-complexity samples while gradually introducing more challenging high-frequency or noise-dominated measurements, thereby mitigating overfitting and accelerating practical inversion. Theoretically, we show that, with an early-stage preference for easy-to-fit measurements, CogGen admits a reduced local sufficient-iteration bound and a smaller cumulative noise-amplification bound within a finite iteration budget, offering a theoretical explanation for improved convergence behavior and image fidelity. Numerical experiments further demonstrate that both CogGen instantiations, CogGen-DIP and CogGen-INR, consistently outperform representative CS-MRI reconstruction techniques, including unsupervised and supervised pipelines.\\
\indent The key contributions of our work are summarized as follows, covering the framework design, scheduling mechanism, theoretical analysis, and experimental validation.
\begin{enumerate}[label=\arabic*., leftmargin=1.5em, labelsep=0.4em,
	itemsep=0pt, topsep=0pt, parsep=0pt, partopsep=0pt]
	\item We propose CogGen, a cognitive-load-inspired FU-DGM framework that reformulates CS-MRI reconstruction as a staged inversion problem and progressively schedules k-space measurement to improve the efficiency and accuracy of untrained generative reconstruction.
	\item We design an MRI-aware SPCL-based k-space scheduling strategy equipped with a dual-threshold sample-weighting criterion, where data-consistency residual thresholding evaluates the fitting reliability of the current generator, whereas k-space radius thresholding defines a stage-wise feasible region to regulate measurement exposure.	
	\item We provide the local and conditional analyses showing that stage-wise measurement weighting can reduce the sufficient-iteration bound and suppress cumulative noise amplification during finite-iteration reconstruction.	
	\item We instantiate CogGen with DIP and INR, and extensive experiments demonstrate that the resulting CogGen-DIP and CogGen-INR outperform representative CS-MRI reconstruction techniques, including both unsupervised and supervised pipelines.
\end{enumerate}

\indent The remainder of this paper is organized as follows. Section \uppercase\expandafter{\romannumeral 2} reviews the preliminaries of CS-MRI reconstruction and classical FU-DGM backbones. Section \uppercase\expandafter{\romannumeral 3} presents the proposed CogGen framework, including the staged inversion formulation and the SPCL-driven k-space scheduling strategy. Section \uppercase\expandafter{\romannumeral 4} reports the experimental settings, comparative results, and ablation studies. Section \uppercase\expandafter{\romannumeral 5} concludes the paper. Finally, the Appendix provides theoretical analyses of CogGen.
\section{Preliminaries}
\subsection{CS-MRI Reconstruction}
A discrete forward observation model for CS-MRI reconstruction can be mathematically expressed as
\begin{equation}
	y = A x + \epsilon,
\end{equation}
where \(x \in \mathbb{C}^N\) denotes the target MR image and \(y \in \mathbb{C}^M\) represents the acquired undersampled k-space measurements. The observation operator \(A: \mathbb{C}^N \rightarrow \mathbb{C}^M\) characterizes the MR acquisition process, typically including coil-sensitivity encoding, Fourier transformation, and k-space undersampling, with M denoting the total number of acquired measurements. Due to the inherent undersampling with \(M \ll N\), the operator \(A\) is rank-deficient or severely ill-posed. The term \(\epsilon\) denotes measurement noise, usually modeled as additive white Gaussian noise (AWGN). Therefore, CS-MRI reconstruction aims to recover a high-quality image \(x\) from the undersampled observation \(y\).

\subsection{Classical FU-DGM: DIP and INR}

\begin{myDef}[Deep Network Prior~\cite{jagatap2019phase, saragadam2024deeptensor}]
	\label{D1}
	{A particular image ${x}$ is deemed to follow an untrained neural network prior (i.e., a fully unsupervised generative prior) if it falls within a set $S$ defined as $S:= \left\{{x|x = f_{\theta}\left(z\right)}\right\}$, where $z$ denotes the underlying latent representation driving the generation.}
\end{myDef}
We build upon two established FU-DGM frameworks, DIP and INR, both of which serve as deep network prior$^{\ref{D1}}$-based image generators with architectural bias and employ the following objective function,
\begin{equation}
	\hat{\theta} = \arg\min_{\theta}
	\parallel Af_{\theta}(z) - y \parallel_2^2,~\hat{x} = f_{\hat{\theta}}(z)
\end{equation}where $f_{\theta}: \mathbb{R}^p \rightarrow \mathbb{C}^{N}$ is a parametric mapping with learnable parameters $\theta$ that generates a complex-valued image. Unlike supervised generative reconstruction which requires external training data, the parameters \(\theta\) are optimized separately for each individual \(y\). DIP injects randomness through $z$ as a latent code, while INR treats $z$ as a coordinate descriptor of the signal domain. As a result, classical FU-DGM formulations, despite yielding visually plausible results, often suffer from suboptimal fine-detail accuracy because prolonged iterative fitting in noisy and ill-posed systems can gradually amplify noise corruption in the reconstruction\cite{heckel2020denoising, buskulic2023convergence, buskulic2024convergence}. Specifically, we have 
\begin{equation}
	x^{(l+1)} = x^{(l)} - \eta A^{\mathrm{H}}\big(Ax^{(l)} - y\big), \quad \eta>0
\end{equation} Substituting \(y = A x^* + \epsilon\) yields 
\begin{equation}
	e^{(l+1)} = \big(I - \eta A^{\mathrm{H}} A\big)e^{(l)} + \eta A^{\mathrm{H}} \epsilon
\end{equation} where \(e^{(l)} = x^{(l)} - x^*\). Let the singular value decomposition (SVD) be \(A = U \Sigma V^{\mathrm{H}}\), with singular values \(\{\sigma_i\}\). Projecting onto the \(i\)-th right singular vector \(v_i\) gives the scalar recursion
\begin{equation}
	e_i^{(l+1)} = (1-\eta \sigma_i^2)\, e_i^{(l)} + \eta \sigma_i\, \epsilon_i
\end{equation}where \(e_i^{(l)} = v_i^{\mathrm{H}} e^{(l)}\), \(\varepsilon_i = u_i^{\mathrm{H}}  \epsilon\). 
Assuming \(0<\eta<2/\sigma_{\max}^2\) so that \(|1-\eta\sigma_i^2|<1\), the steady-state limit satisfies $e_i^{(\infty)} \approx \frac{\epsilon_i}{\sigma_i}$. In CS-MRI, such small-singular-value directions are typically associated with weakly sampled or high-frequency components, where measurement noise and aliasing artifacts are more easily imprinted into the reconstruction. Furthermore, from an optimization perspective, the ill-conditioning of $A$, characterized by a large condition number $\kappa(A) = \sigma_{\max}/\sigma_{\min}$, also slows down iterative reconstruction, i.e., the components corresponding to extremely small singular values exhibit a convergence factor $|1-\eta\sigma_i^2|$ very close to 1. 
\section{Proposed Framework: CogGen}
\begin{figure}[t]
	\centering
	\includegraphics[width=0.7\linewidth]{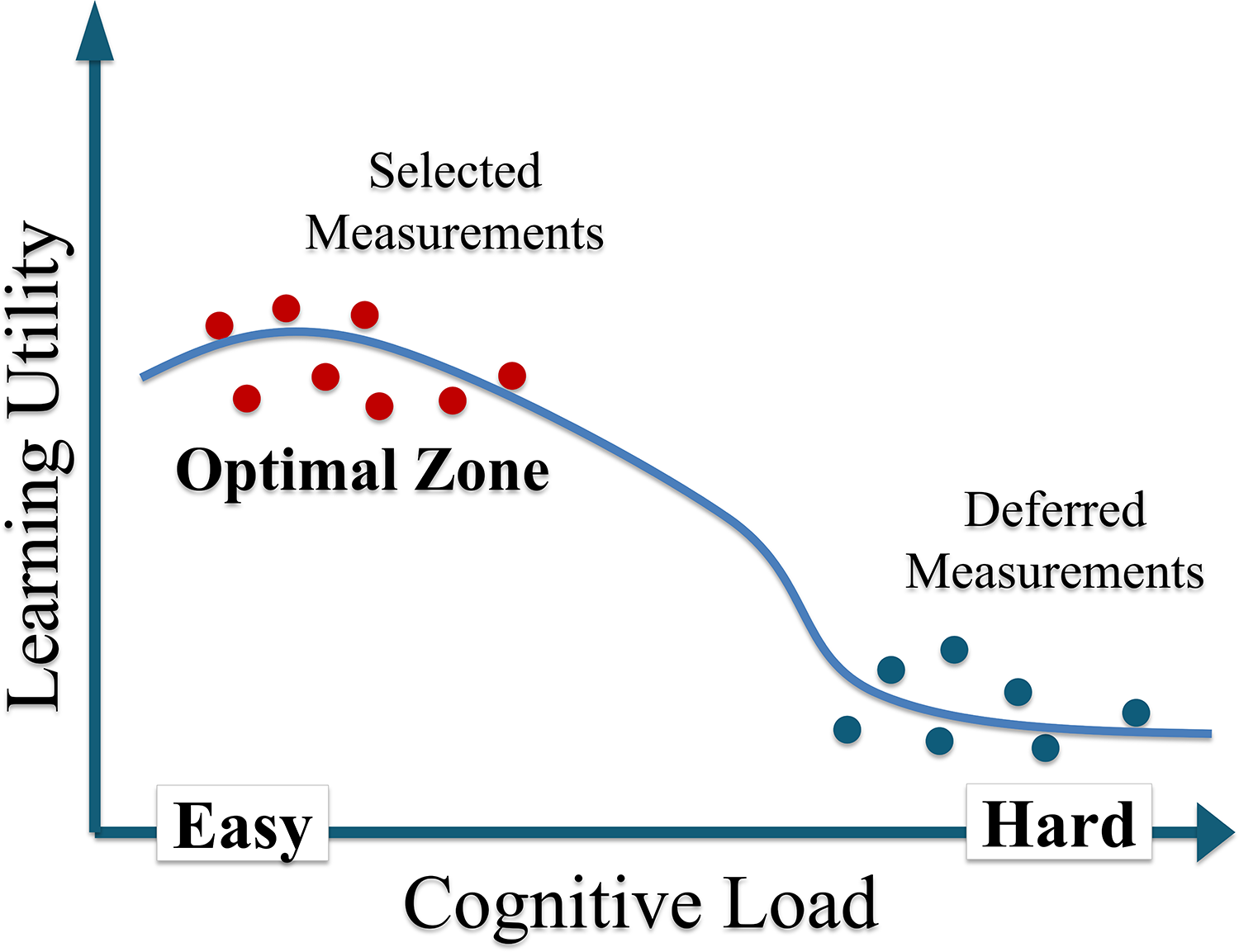}  
	\caption{Cognitive-load view of CogGen: measurements with lower load exhibit higher learning utility and are therefore selected earlier, whereas high-load measurements with reduced suitability are deferred to the further stages.}	
\end{figure}
In cognitive psychology, cognitive-load theory emphasizes that optimal learning utility is achieved by dynamically aligning task demands with the finite processing capacity of the learner to avoid information overload. Inspired by this principle, we propose CogGen, a cognitive-load-inspired FU-DGM framework that reformulates CS-MRI reconstruction as a staged inversion problem rather than a uniform iterative fitting task. In the context of CS-MRI, the cognitive load can be interpreted as the effective inversion burden imposed by heterogeneous k-space measurements, including sample-specific difficulty and noise-induced interference. Instead of estimating the target image by matching all undersampled k-space measurements uniformly throughout the optimization process, CogGen performs progressive scheduling over k-space measurements (As illustrated in Fig.1) and decomposes the original ill-posed inverse problem into a sequence of progressively refined subproblems, each associated with a controlled subset of measurements. This easy-to-hard inversion design~\cite{liu2025uhd}, which mirrors human cognitive learning patterns, enables the model to first solve structurally simpler and more stable inversion objectives, and then gradually incorporate increasingly difficult components, thus matching the fitting burden to the evolving representation capacity of the generator while improving reconstruction quality within a finite iteration budget, and mitigating semi-convergence by delaying the fitting of noise-dominated measurements. 

\subsection{SPCL-based Scheduling Principle}
Mathematically, progressive scheduling can be formulated as a sample-weighting problem ~\cite{zhou2023investigating,shu2023cmw,wu2025data} defined over the k-space measurements. Conventional uniform fitting assigns equal weights to all measurements, forcing the generator to match the acquired data with equal strength. However, this ignores the intrinsic heterogeneity of k-space: frequency-domain coefficients near the center encode dominant low-frequency anatomical structures with high SNR and stability, whereas peripheral high-frequency coefficients are highly vulnerable to undersampling-induced ill-posedness and noise contamination. Therefore, a scheduling scheme requires non-uniform, stage-dependent weighting. In this work, we adopt an MRI-aware dual-threshold weighting criterion-based SPCL strategy, which integrates curriculum learning~\cite{ref48} and self-paced learning~\cite{ref28} by combining a predefined easy-to-hard prior with adaptive sample selection according to the current learning state. Consequently, sample importance under soft-weighting or hard-selection regimes is assigned in a difficulty-aware and state-adaptive manner, enabling the generator to fit simpler and more reliable structures before gradually incorporating more challenging measurements.

\subsection{Detailed Formulation}
We first define a stage sequence over k-space,
\begin{equation}
	C_1 \subset C_2 \subset \cdots \subset C_n \subset \cdots \subset C_T
\end{equation}where $C_n$ denotes the set of acquired k-space measurements included in the $n$-th stage, and each measurement is treated as a schedulable sample. Thus, $C_n$ specifies the measurements that are progressively exposed to the generator at stage $n$, rather than forcing all acquired measurements to participate with equal strength from the beginning. In general, the entire set of k-space measurements is covered only at the final stage. Accordingly, conditioned on the stage-wise weights \(s_{C_n}\in[0,1]^{M}\), a regularized optimization problem is formulated over the network parameters \(\theta_{C_n}\),
\begin{equation}
	\begin{aligned}
		\hat{\theta}_{C_n} = ~&\underset{\theta_{C_n}}{\arg\min}\;
		{\left\|s_{C_n} \circ \left(A f_{\theta_{C_n}}(z) - y\right) \right\|_2}/{\left\| s_{C_n} \circ y \right\|_2} \\
		&+ \gamma \left(\left\|\nabla_h f_{\theta_{C_n}}(z)\right\|_1 + \left\|\nabla_v f_{\theta_{C_n}}(z)\right\|_1 \right)\\
	\end{aligned}
\end{equation}where the first term enforces normalized weighted data-fidelity and the second applies anisotropic total-variation (TV) constraint. \(\circ\) denotes the Hadamard product. \(\nabla_h\) and \(\nabla_v\) denote the horizontal and vertical discrete gradient operators, respectively. The hyperparameter \(\gamma>0\) balances the data fidelity against the regularizer.\\
\indent Consequently, we first define the stage-wise weighting vector $s_{C_n}$ with the following closed-form rule. Here, the larger weight $w$ emphasizes reliable measurements, whereas 1-$w$ retains the remaining measurements with reduced influence; in particular, $w$ = 1 corresponds to hard selection, while $w<1$ yields soft weighting.
\begin{equation}
	(s_{C_n})_i^*
	=
	w\mathbf{1}_{{\mathcal{I}}_{C_n}}(i)
	+
	(1-w)\mathbf{1}_{{\mathcal{I}}_{C_n}^c}(i),~ \frac{1}{2}<w\leq 1
\end{equation}
with
\begin{equation}
	{\mathcal{I}}_{C_n}
	=
	\left\{
	i\in t_{C_n}
	\;\middle|\;
	|\left(A f_{\theta_{C_n}}(z)\right)_i-y_i|/|y_i|<\lambda_{C_n}
	\right\}
\end{equation}
where $\mathbf{1}_{{\mathcal{I}}_{C_n}}$ is the indicator function of ${\mathcal{I}}_{C_n}$, yielding 1 if $i\in{\mathcal{I}}_{C_n}$ and 0 otherwise, and ${\mathcal{I}}_{C_n}^{c}$ denotes its set-theoretic complement. The scalar $\lambda_{C_n}$ serves as an adaptive threshold for data-consistency residual thresholding. $t_{C_n}$ designates the feasible region, formally defined as the following index set,
\begin{equation}
	t_{C_n}
	=
	\left\{
	i 
	\;\middle|\;
	\|\mathbf{k}_i-\mathbf{k}_{\mathrm{center}}\|_2 < r_{C_n}
	\right\}
\end{equation}where $\mathbf{k}_i \in \mathbb{R}^2$ denotes the coordinate vector of the $i$-th k-space measurement, and $\mathbf{k}_{\text{center}} \in \mathbb{R}^2$ is the frequency-domain origin. \(r_{C_n}\) is a dynamic k-space radius that expands progressively across stages. Therefore, ${\mathcal{I}}_{C_n}$ contains only those measurements that simultaneously satisfy the feasible-region constraint $i\in t_{C_n}$ and the residual criterion, ensuring that a sample is strongly weighted only when it is both scheduled to be exposed and currently reliable to fit.\\
\indent After each scheduling stage, the parameters $\lambda_{C_n}$ and $r_{C_n}$ are subjected to a deterministic update rule,
\begin{equation}
	\begin{aligned}
\lambda_{C_n}& = \lambda_{C_1}\left(\frac{\lambda_{C_T}}{\lambda_{C_1}}\right)^{\frac{n-1}{T-1}}\\
r_{C_n} &= r_{C_1}+\frac{n-1}{T-1}\left(r_{C_T}-r_{C_1}\right)
	\end{aligned}
\end{equation}where $\lambda_{C_1}$ and $r_{C_1}$ were initialized with small values and progressively enlarged across stages. Conditioned on the resulting weights, the network parameters are updated for $L(n)$ inner iterations as
\begin{equation}
	\begin{split}
		\theta_{C_n}^{(l+1)}
		&=
		\theta_{C_n}^{(l)}
		-\tau \nabla_{\theta_{C_n}^{(l)}}
		\bigl(
		\|s_{C_n}\odot(Af_{\theta_{C_n}}(z)-y)\|_2/\|s_{C_n}\odot y\|_2
		\\
		&
		+\gamma\bigl(
		\|\nabla_h f_{\theta_{C_n}}(z)\|_1+\|\nabla_v f_{\theta_{C_n}}(z)\|_1
		\bigr)
		\bigl)
	\end{split}
	\raisetag{10pt}
\end{equation}
where $\tau>0$ is the step size. Algorithm~\ref{alg:CogGen} provides a summary.
\begin{algorithm}[t]
	\caption{CogGen}
	\label{alg:CogGen}
	\renewcommand{\algorithmicrequire}{\textbf{Input:}}
	\renewcommand{\algorithmicensure}{\textbf{Output:}}
	\begin{algorithmic}[1]
		\REQUIRE $A$, $y$, $w$, $z$, $\lambda_{C_n}$, $r_{C_n}$, $\Delta\lambda$, $\Delta r$, $T$, $L$
		
		\FOR{$n = 1$ to $T$}
		\STATE Update $s_{C_n}$ according to Eq.~(8).
		\FOR{$l = 1$ to $L(n)$}
		\STATE Update $\theta_{C_n}$ according to Eq.~(12).
		\ENDFOR
		
		\STATE Update $\lambda_{C_n}$, $r_{C_n}$ according to Eq.~(11).
		\ENDFOR
		
		\ENSURE $\hat{x} = f_{\theta}(z)$
	\end{algorithmic}
\end{algorithm}

\section{Experiments and Results}

\subsection{Experimental Setup}

To assess the performance of our CogGen framework, the retrospective CS-MRI experiments were conducted on the following four publicly available in-vivo human datasets, for which ethical approval and informed-consent procedures were handled according to the original data-acquisition protocols of the corresponding data sources:

\noindent\textbf{Data$\#1$}~\cite{ref41}: 3D brain data from healthy subjects were acquired using a 12-channel coil and a 3D T2 CUBE sequence (readout (RO) × phase-encoding (PE) × Slice: $256 \times 232 \times 208$). A representative slice was retrospectively undersampled with a 2D variable-density (VD) random pattern at an acceleration factor (AF) of 8.

\noindent\textbf{Data$\#2$}~\cite{ref42}: 2D knee data were acquired on a 3T Siemens scanner using a TSE sequence and a 15-channel coil (RO × PE × Slice: $320 \times 320 \times 227$). A representative slice was retrospectively undersampled with a 1D VD random pattern at an AF of 6.

\noindent\textbf{Data$\#3$}~\cite{zhu2019}: Brain data were acquired on a 3T Siemens scanner using a gradient-echo sequence and a 12-channel head coil array (RO × PE × Slice: $256 \times 256 \times 20$). A 1D VD random undersampling pattern at an AF of 5 was retrospectively applied.

\noindent\textbf{Data$\#4$}~\cite{peng2015incorporating}: An independent multicontrast MRI dataset was acquired on a Siemens Trio Tim 3.0T system using a TSE sequence (RO $\times$ PE $\times$ Contrast: $384 \times 324 \times 3$). A 2D VD random undersampling pattern at an AF of 10 was used during acquisition.

\indent We provide two instantiations of the CogGen framework, namely CogGen-DIP and CogGen-INR, benchmarking them against several SOTA CS-MRI reconstruction techniques including DIP-TV~\cite{ref43}, BM3D-FISTA~\cite{ref46}, DISCUS~\cite{ref47}, SSDU~\cite{yaman2020self}, aSeq-DIP~\cite{ref45}, Hash-INR-$\ell_2$-$\ell_1$~\cite{xu2025self}, PDAC~\cite{wang2024progressive} and MoDL \cite{ref41}. Except for SSDU, PDAC, and MoDL, all remaining competing methods operate in a training-free manner. PDAC and MoDL were trained using approximately 300$\sim$400 external training samples, whereas SSDU adopted a scan-specific self-supervised data-splitting strategy without requiring external paired training data. Notably, PDAC is included as a mechanistically relevant white-box baseline because it also performs progressive reconstruction. The DIP-based baselines were implemented using a deep U-Net variant with a 5-level encoder-decoder, 128 feature channels, and skip connections between corresponding encoder and decoder stages. Conversely, the INR counterpart adopted a hash-encoded multilayer perceptron with sinusoidal activations, consisting of 8 hidden layers with 256 neurons each, and further incorporated Fourier feature mappings to better represent high-frequency image details, optimized with a hybrid $\ell_2$-$\ell_1$ data-consistency loss. All deep learning-based models were implemented in PyTorch and executed on an NVIDIA RTX A6000 GPU, and were optimized using Adam with a learning rate of $1\times 10^{-4}$. Moreover, for all methods, we systematically tuned the model and regularization hyperparameters to ensure a fair and controlled comparison. For CogGen, we set $T=5$ and $L=[1000,1000,2000,2000,10000]$; In addition, $\lambda_{C_1}$ and $r_{C_1}$ are set within the ranges of $[1\times10^{-6}, 1\times10^{-5}]$ and $[60,80]$, respectively. At the final stage, $\lambda_{C_T}$ is set within $[10,15]$, and $r_{C_T}$ is chosen as $r_{C_T}=\max_{i\in\Omega}|\mathbf{k}_i-\mathbf{k}_{\mathrm{center}}|_2$. Quantitative evaluation was conducted using two metrics, the relative $\ell_2$-norm error (RLNE) and the peak signal-to-noise ratio (PSNR), both computed within a region of interest (ROI) that encompassed the brain and knee anatomical structures while excluding the background, and whose explicit definitions are given below:
\begin{align}
	\text{RLNE}_{\text{ROI}} &= \frac{\| x_{\text{ROI}} - \hat{x}_{\text{ROI}} \|_2}{\| x_{\text{ROI}}\|_2}\times 100\% \\
	\text{PSNR}_{\text{ROI}} &= 20 \log_{10} \left( \frac{\max(x_{\text{ROI}})}{\sqrt{\frac{1}{N} \| x_{\text{ROI}} - \hat{x}_{\text{ROI}} \|_2^2}} \right)
\end{align}
\begin{figure*}[t]
	\centering
	\includegraphics[width=1\linewidth]{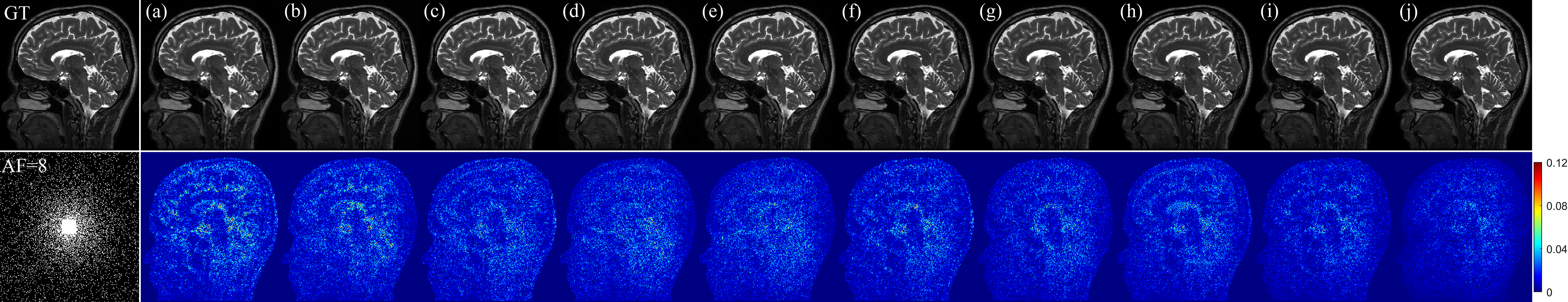}  
	\caption{Reconstruction results on Data$\#1$ at AF = 8. (a)-(j): DIP-TV, BM3D-FISTA, DISCUS, SSDU, aSeq-DIP, Hash-INR-$\ell_2$-$\ell_1$, PDAC, MoDL, CogGen-DIP and CogGen-INR. The corresponding error maps are presented in the bottom row, alongside the GT and downsampling mask displayed at the far left.}
	\label{fig_2}
\end{figure*}
\begin{figure*}[t]
	\centering
	\includegraphics[width=1\linewidth]{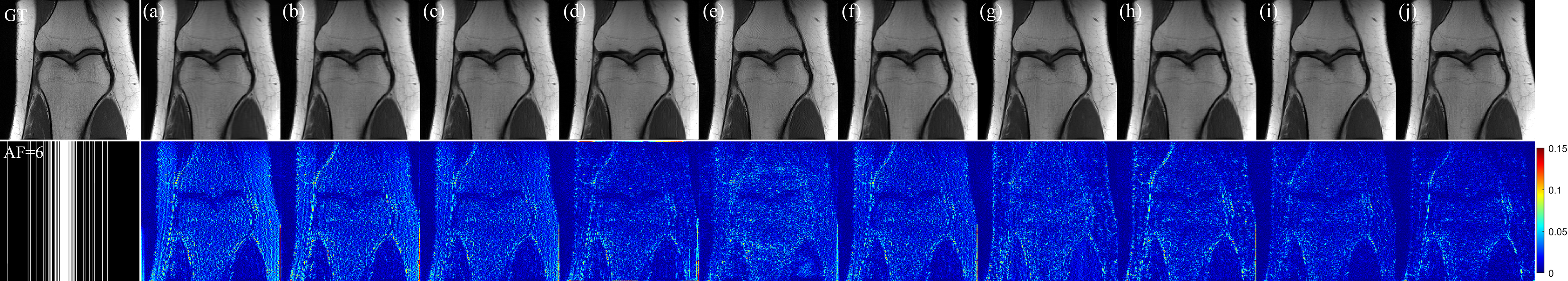}  
	\caption{Reconstruction results on Data$\#2$ at AF = 6. (a)-(j): DIP-TV, BM3D-FISTA, DISCUS, SSDU, aSeq-DIP, Hash-INR-$\ell_2$-$\ell_1$, PDAC, MoDL, CogGen-DIP and CogGen-INR. The corresponding error maps are presented in the bottom row, alongside the GT and downsampling mask displayed at the far left.}
	\label{fig_2}
\end{figure*}
\begin{figure}[h]
	\centering
	\includegraphics[width=0.96\linewidth]{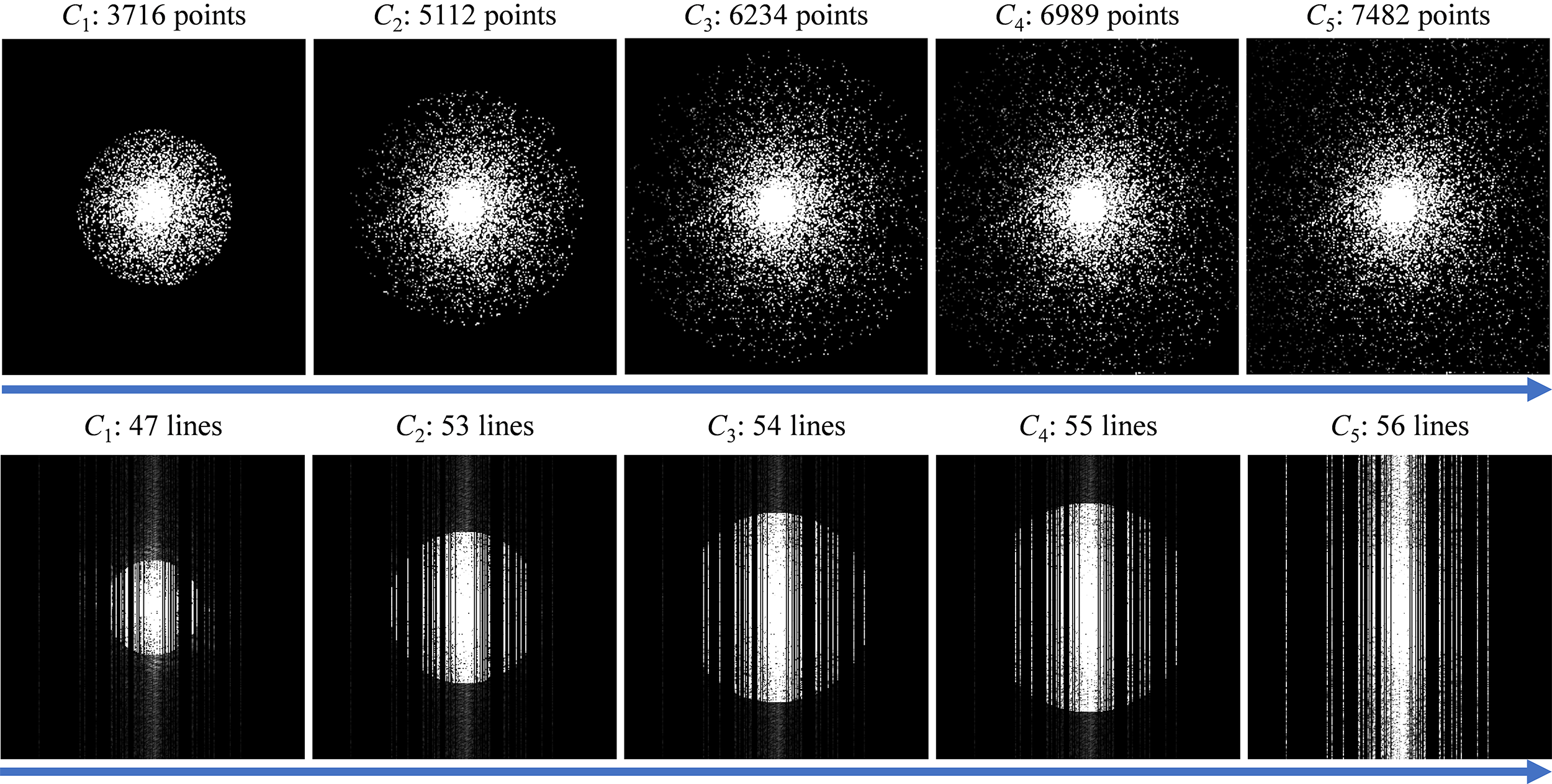} 
	\caption{Evolution of sample selection across five stages ($C_1$-$C_5$), showing the expansion from central k-space to the full acquired pattern. Top: Data$\#1$ (2D, $w=1$). Bottom: Data$\#2$ (1D, $w=0.8$). Here, $C_5$ denotes the final stage, which contains the original undersampled k-space measurements.}
	\label{fig_1}
\end{figure}
\begin{table}[h]
	\centering
	\caption{
		Quantitative results ($\text{RLNE}_{\text{ROI}}$ and $\text{PSNR}_{\text{ROI}}$) on Data$\#1$ (2D, AF = 8) and Data $\#2$ (1D, AF = 6). The top two performing entries are bolded.}
	\resizebox{1\linewidth}{!}{
		\renewcommand{\arraystretch}{1}  
		\begin{tabular}{lcccc}
			\hline
			\multirow{2}{*}{Methods} & \multicolumn{2}{c}{Data$\#1$ (2D, AF = 8)} & \multicolumn{2}{c}{Data$\#2$ (1D, AF = 6)} \\
			\cmidrule(lr){2-3} \cmidrule(lr){4-5} 
			& RLNE$_{\text{ROI}}$(\%) & PSNR$_{\text{ROI}}$(dB) & RLNE$_{\text{ROI}}$(\%) & PSNR$_{\text{ROI}}$(dB) \\
			\hline
			DIP-TV & 9.84 & 38.17 & 6.87 & 30.65 \\
			BM3D-FISTA & 9.60 & 38.45 & 6.28 & 31.42 \\
			DISCUS & 8.59 & 38.91 & 5.66 & 32.32 \\
			SSDU & 8.52 & 38.91 & 4.83 & 33.79 \\
			aSeq-DIP & 8.41 & 39.24 & 5.42 & 32.77 \\
			Hash-INR-$\ell_2$-$\ell_1$ & 8.17 & 39.43 & 5.30 & 32.90 \\
			PDAC & 7.56 & 40.45 & 4.76 & 33.84 \\
			MoDL & 7.89 & 39.97 & 4.65 & 34.03 \\
			\textbf{CogGen-DIP} & \textbf{6.70} & \textbf{41.50} & \textbf{3.89} &
			\textbf{35.58} \\
			\textbf{CogGen-INR} & \textbf{6.06} & \textbf{42.42} & \textbf{3.45} & \textbf{36.63} \\
			\hline
		\end{tabular}
	}
	\label{tab3}
\end{table}
\label{sec:ablation}
\begin{figure}[!t]
	\centering
	\includegraphics[width=1.02\linewidth]{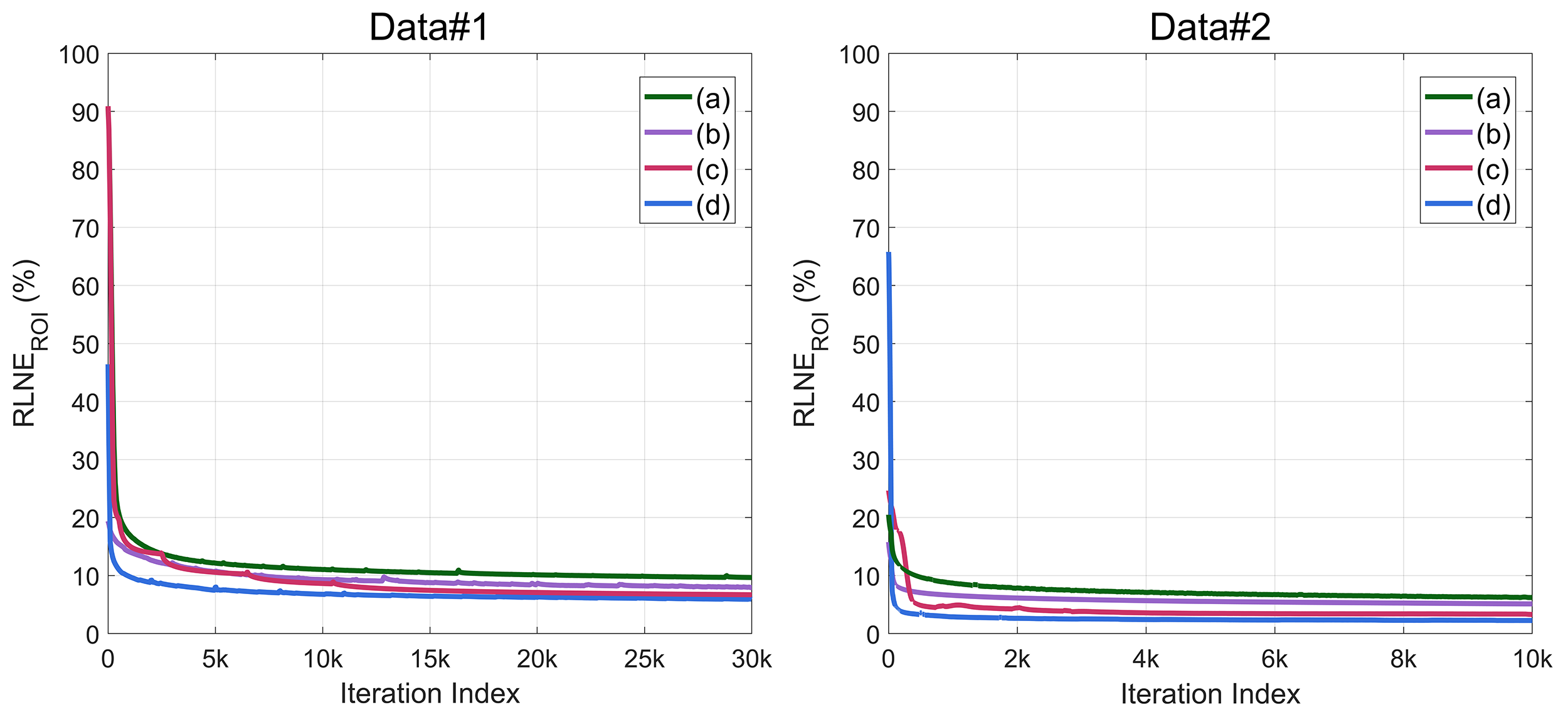}
	\caption{Performance gains from CogGen on Data$\#1$ (2D, AF = 8) and Data$\#2$ (1D, AF = 6). Integrating CogGen into DIP- and INR-based baselines notably boosts accuracy and convergence under the same iteration budget. (a)-(d): DIP-TV, Hash-INR-$\ell_2$-$\ell_1$, CogGen-DIP, and CogGen-INR.}
\end{figure} 
\subsection{Reconstruction Performances}
We present the reconstruction results in Fig.~2 (Data$\#1$, AF = 8) and Fig.~3 (Data$\#2$, AF = 6). It is evident that DIP-TV and BM3D-FISTA yield inferior reconstructions, with noticeably blurred structures and oversmoothed details, whereas DISCUS, SSDU, aSeq-DIP and Hash-INR-$\ell_2$-$\ell_1$ achieve broadly comparable visual quality by effectively suppressing noise-like artifacts while preserving delicate anatomical patterns. Notably, SSDU exhibits pronounced performance degradation under the more challenging AF = 8 setting. Meanwhile, PDAC and MoDL, functioning as the fully supervised methods, 
\begin{figure*}[t]
	\centering
	\includegraphics[width=0.76\linewidth]{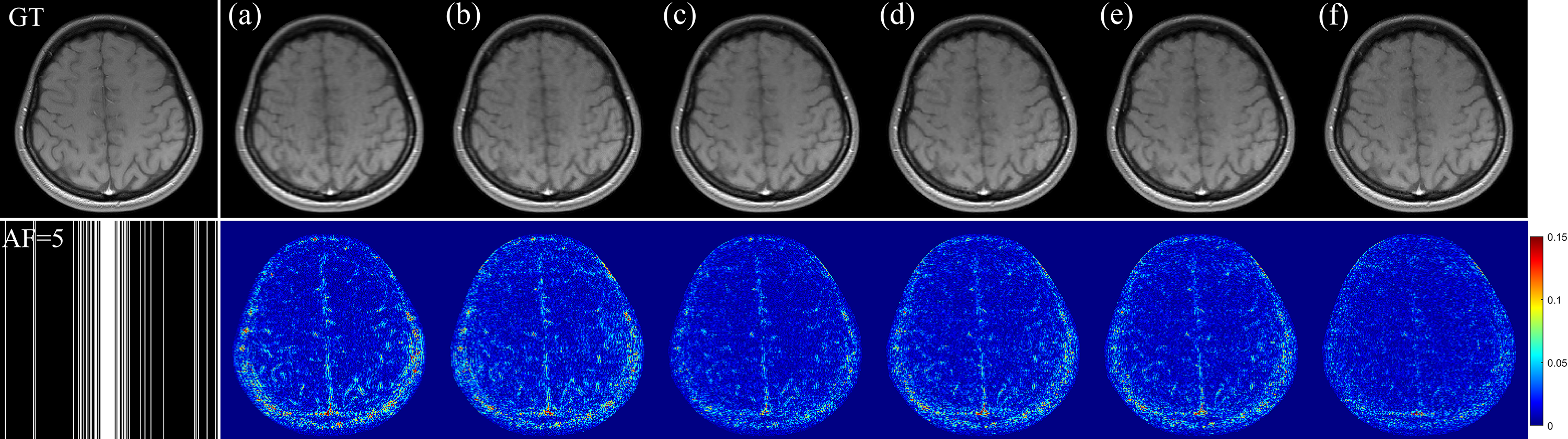}
	\caption{Comparison between MRI-Aware and Random Scheduling on Data$\#3$ at an AF of 5. (a)-(c): DIP-TV, SRS-DIP, CogGen-DIP; (d)-(f): Hash-INR-$\ell_2$-$\ell_1$, SRS-INR, CogGen-INR. The corresponding error maps are presented in the bottom row, alongside the GT and downsampling mask displayed at the far left.}
	\label{fig_6}
\end{figure*}
\begin{table}[!t]
	\centering
	\caption{Quantitative evaluation of MRI-aware vs. random scheduling on Data$\#3$ at AF = 5.}
	\resizebox{0.65\linewidth}{!}{
		\begin{tabular}{lcc}
			\toprule
			\multicolumn{1}{c}{\multirow{2}{*}{Methods}} & \multicolumn{2}{c}{Data$\#3$ (1D, AF = 5)} \\
			\cmidrule(lr){2-3}
			& RLNE\textsubscript{ROI}(\%) & PSNR\textsubscript{ROI}(dB) \\
			\midrule
			DIP-TV            & 8.13          & 37.58          \\
			SRS-DIP      & 7.96          & 37.76          \\
			\textbf{CogGen-DIP}& \textbf{6.04} & \textbf{40.13} \\
			Hash-INR-$\ell_2$-$\ell_1$   & 7.01          & 38.86          \\
			SRS-INR      & 6.95          & 38.95          \\
			\textbf{CogGen-INR}& \textbf{5.77} & \textbf{40.57} \\
			\bottomrule
		\end{tabular}
	}	
\end{table}consistently outperform the remaining non-supervised competing baselines. Empowered by the CogGen scheduling procedure illustrated in Fig.~4, with $w=1$ for Data$\#1$ and $w=0.8$ for Data$\#2$, our proposed CogGen-DIP and CogGen-INR yield reconstructions that most closely match the GTs, with higher fidelity in recovering intricate textures and sharp structural boundaries. This improvement is also evident from the corresponding error maps, where CogGen produces weaker residual artifacts around anatomical edges and fine structures, indicating that progressive k-space scheduling helps suppress noise imprint while preserving diagnostically relevant details. These visual gains are reflected in Table.\uppercase\expandafter{\romannumeral 1}, where both variants achieve the best performances across the primary evaluation metrics.
\subsection{Discussion}
\subsubsection{CogGen superiority} In Fig.~5, we plot the reconstruction curves of $\text{RLNE}_{\text{ROI}}$ and $\text{PSNR}_{\text{ROI}}$ for DIP- and INR-based baselines (DIP-TV and Hash-INR-$\ell_2$-$\ell_1$) on Data$\#1$ and Data$\#2$, with and without the CogGen strategy. The results clearly reveal that CogGen consistently boosts both reconstruction accuracy and convergence efficiency across both backbones. Notably, to achieve comparable fidelity, CogGen requires substantially fewer iterations, highlighting a distinct low-latency advantage. Compared with DIP-TV, the Hash-INR-$\ell_2$-$\ell_1$-based instantiation enjoys a more pronounced gain, suggesting that the global functional parameterization of INR may benefit more directly from the easy-to-hard measurement pacing. Furthermore, during the early optimization stages, reconstructing from a dynamically constrained data subset yields noticeably higher precision than naively fitting the entire acquired measurement set from the start. This reveals that the capacity of FU-DGMs can be better exploited when measurement exposure is aligned with the current learning state. In other words, CogGen keeps the backbone generator unchanged and instead controls which measurements are emphasized during optimization. Structurally informative measurements are fitted first, while more difficult or noise-sensitive components are gradually introduced in later stages.\\
\indent To further evaluate the superiority of CogGen, we conduct an additional ablation study of MRI-aware scheduling on Data$\#3$ at AF = 5, as shown in Fig.~6. Specifically, we construct two staged random-scheduling variants based on DIP-TV and Hash-INR-$\ell_2$-$\ell_1$, namely SRS-DIP and SRS-INR. At each stage, the two variants activate exactly the same number of k-space measurements as their corresponding CogGen counterparts, but randomly select the participating samples instead of using the proposed dual-threshold criterion. The results indicate an interesting observation: even random progressive exposure slightly improves reconstruction compared with fitting the entire acquired measurement set from the beginning. This suggests that gradually enlarging the active measurement set may itself provide a mild regularizing effect for FU-DGM optimization. Nevertheless, CogGen consistently yields cleaner reconstructions and lower errors than its random-scheduling counterparts, demonstrating that MRI-aware scheduling plays a critical role in guiding a more accurate reconstruction trajectory. The quantitative results are reported in Table.\uppercase\expandafter{\romannumeral 2}.
\subsubsection{The influence of stage size} 
To isolate the effect of the stage size while avoiding excessive computational cost, we perform this controlled evaluation using CogGen-INR on Data$\#4$ at AF = 10. As shown in Fig.~7, the quantitative results exhibit a clear dependence on the stage size, indicating the presence of an appropriate stage count at which the generative capacity of INR can be more effectively exploited. This behavior is consistent with the easy-to-hard scheduling rationale: a properly paced stage facilitates stable inversion, whereas an overly coarse schedule with too few stages, e.g., $T<3$, may expose difficult measurements prematurely. In contrast, an overly fragmented schedule with too many stages, e.g., $T>5$, may allocate too few iterations to each stage and slow down the incorporation of useful measurements. Both cases can hinder the staged inversion process and eventually degrade reconstruction quality.
\subsubsection{Single-Thresholding vs. Dual-Thresholding} 
To validate the necessity of this dual-thresholding design, we conduct an component-wise experiment on Data$\#4$ at AF = 10, using an identical INR architecture and the same total iteration budget. As illustrated in Fig.~8, removing the weighting design reduces CogGen to the standard Hash-INR-$\ell_2$-$\ell_1$ baseline and leads to visibly degraded reconstructions. While applying either weighting strategy alone yields clear accuracy gains, each single-thresholding variant has inherent limitations. When only the radius-based constraint is used, difficult samples within the current feasible region may still be strongly weighted before they become reliably fitable, causing suboptimal updates before the optimization has sufficiently stabilized. In contrast, relying solely on the data-consistency residual threshold rule may penalize informative high-frequency components at later stages because it lacks frequency-structured guidance, resulting in blurred textures or localized artifacts. Combining the two modes achieves the best overall performance, improving detail recovery while preserving structural consistency, as highlighted by the red boxes. The quantitative results in Table.\uppercase\expandafter{\romannumeral 3} further support the effectiveness of the dual-thresholding strategy across all metrics.
\begin{figure}[t]
	\centering
	\includegraphics[width=0.72\linewidth]{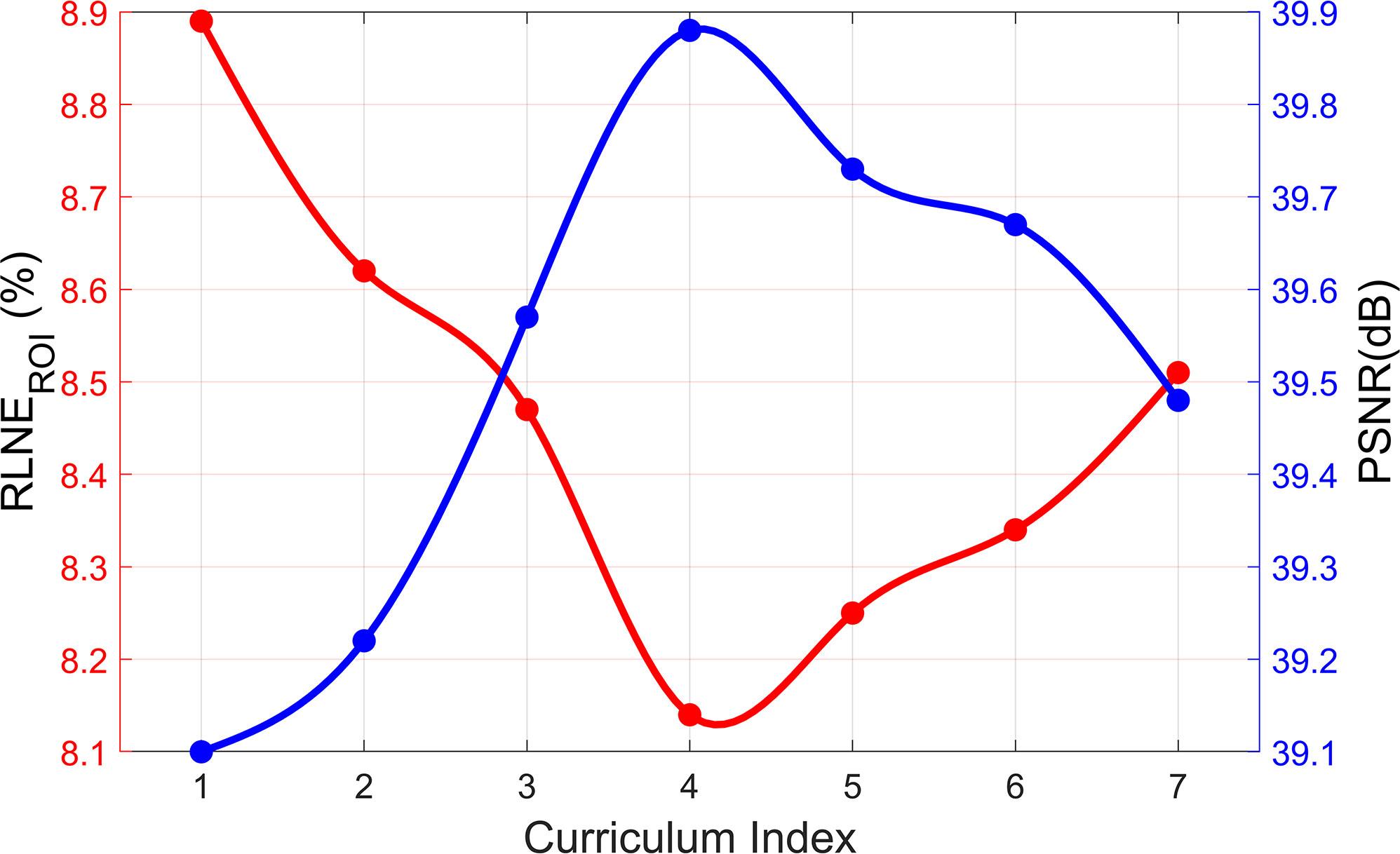}
	\caption{Influence of stage count on Data$\#4$ at AF = 10. Performance peaks at $C_4$ and decreases when additional stages are introduced, highlighting the importance of appropriate stage-count selection for reconstruction performance.}
	\label{fig_5}
\end{figure}
\begin{figure}[t]
	\centering
	\includegraphics[width=0.95\linewidth]{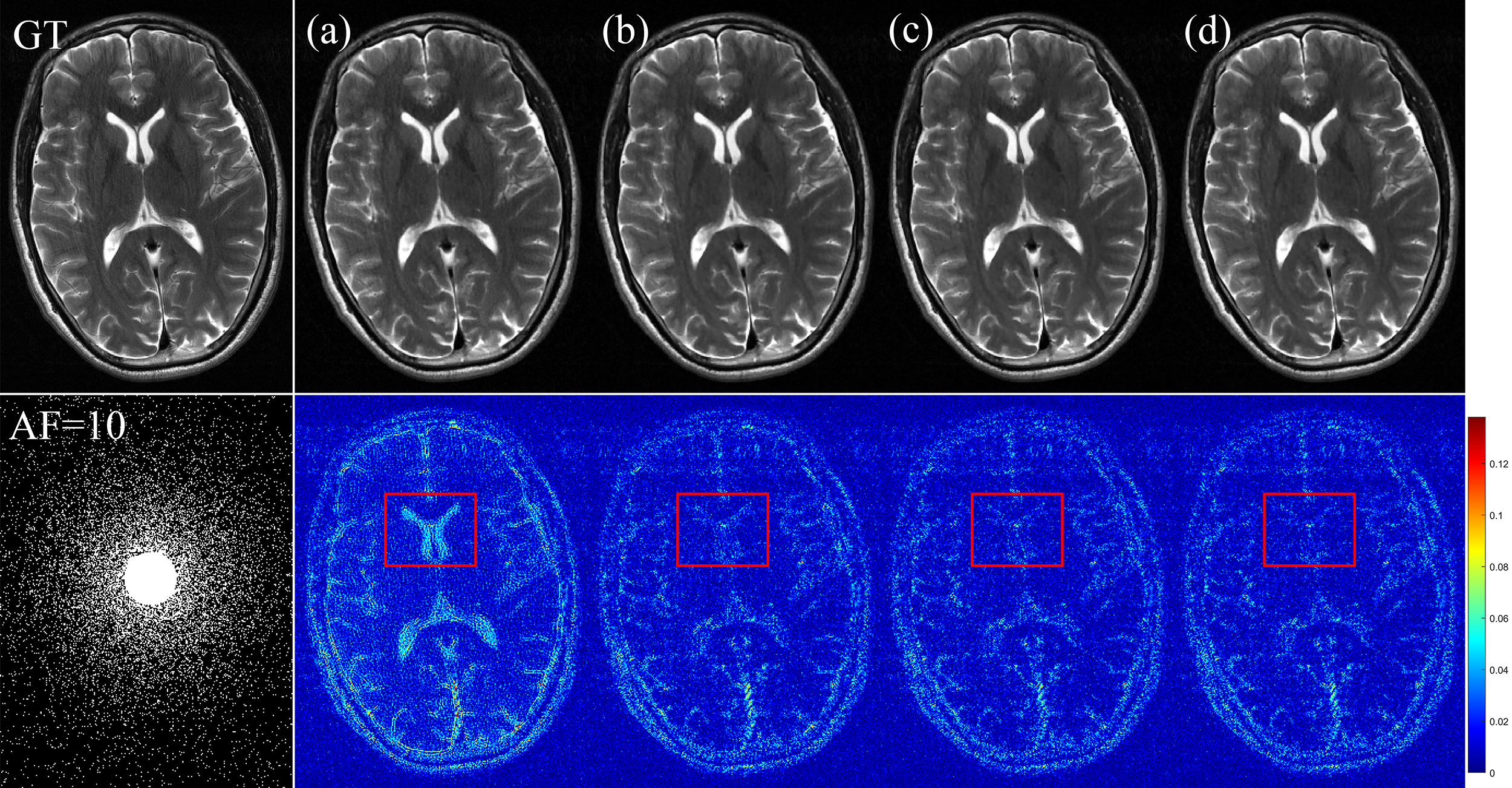}
	\caption{Single-mode vs. dual-mode thresholding (Data$\#4$, AF = 10). Dual-mode scheduling enhances detail recovery and reconstruction fidelity. (a-d): Hash-INR-$\ell_2$-$\ell_1$, CogGen$_{\rm{CL}}$-INR, CogGen$_{\rm{SPL}}$-INR and CogGen-INR.}
	\label{fig_7}
\end{figure}
\begin{table}[h]
	\centering
	\caption{Quantitative comparison for two thresholding schemes on Data$\#4$ (AF = 10).}
	\resizebox{0.8\linewidth}{!}{
		\begin{tabular}{lcc}
			\toprule
			\multicolumn{1}{c}{\multirow{2}{*}{Methods}} & \multicolumn{2}{c}{Data$\#4$ (2D, AF = 10)} \\
			\cmidrule(lr){2-3}
			& RLNE\textsubscript{ROI}(\%) & PSNR\textsubscript{ROI}(dB) \\
			\midrule
			Hash-INR-$\ell_2$-$\ell_1$ & 8.89 & 39.10 \\
			CogGen$_{\rm{CL}}$-INR & 8.41 & 39.59 \\
			CogGen$_{\rm{SPL}}$-INR & 8.29 & 39.71 \\
			\textbf{CogGen-INR} & \textbf{8.14} & \textbf{39.88} \\
			\bottomrule
		\end{tabular}
	}	
	\label{tab:ablation}
\end{table}
\section{Conclusions}
We introduce CogGen, a novel FU-DGM framework that employs a simple yet effective strategy to improve both reconstruction accuracy and convergence efficiency by reformulating conventional uniform k-space fitting as a staged inversion process. The framework is instantiated in two distinct forms, CogGen-DIP and CogGen-INR, whose effectiveness and generality are supported by numerical experiments and theoretical analyses. In future work, we will explore the potential submodular structure of k-space measurements with respect to image-quality assessment to enable principled, adaptive sample selection. Moreover, incorporating submodular optimization strategies \cite{yang2021learning,el2022data}, such as vertex-cover-inspired selection \cite{shi2023greedy,shi2025weak}, may further tighten performance bounds while improving theoretical interpretability.

\section{Appendix}
\setcounter{equation}{0}
\renewcommand{\theequation}{A.\arabic{equation}}
\setcounter{lemma}{0}
\renewcommand{\thelemma}{A.\arabic{lemma}}
\setcounter{theorem}{0}
\renewcommand{\thetheorem}{A.\arabic{theorem}}
This appendix provides the local and conditional analyses: First, we show that, under an early-stage preference for reliably fitted measurements, the CogGen objective admits a reduced local sufficient-iteration bound compared with uniform fitting. Second, we show that early-stage weighting gives a smaller cumulative noise-amplification bound. These analyses explain the observed convergence and accuracy advantages.
\subsection{Local Convergence Behavior of CogGen}
To isolate the role of stage-wise measurement weighting, we analyze the local behavior of the non-squared weighted data-fidelity term used in CogGen. The TV regularizer and the normalization factor are omitted or absorbed into constants, since the present analysis focuses on how the stage-wise measurement operator affects the local optimization trajectory. For a fixed stage $C_n$, consider
\begin{equation}
	\mathcal{L}_{C_n}(\theta)
	=
	\left\|
	S_{C_n}
	\big(Af_{\theta}(z)-y\big)
	\right\|_2,
	\label{eq:A1}
\end{equation}
where $S_{C_n}=\mathrm{diag}(s_{C_n})$ is the diagonal stage-wise weighting operator.
Following a standard local linearization (NTK style) approximation~\cite{golikov2022neural}, around a reference iterate $\theta^{(j)}$, the generator is locally linearized as
\begin{equation}
	f_{\theta}(z)
	\approx
	f_{\theta^{(j)}}(z)
	+
	J^{(j)}
	\big(\theta-\theta^{(j)}\big),
	\label{eq:A2}
\end{equation}
where $J^{(j)}=\nabla_{\theta} f_{\theta^{(j)}}(z)$. Substituting \eqref{eq:A2} into \eqref{eq:A1} gives the weighted local surrogate
\begin{equation}
	\mathcal{L}_{C_n}(\theta)
	\approx
	\left\|
	S_{C_n}
	\Big(
	AJ^{(j)}
	(\theta-\theta^{(j)})
	-
	d^{(j)}
	\Big)
	\right\|_2,
	\label{eq:A3}
\end{equation}
where
\begin{equation}
	d^{(j)}
	=
	y-Af_{\theta^{(j)}}(z).
	\label{eq:A4}
\end{equation}

For brevity, define
\begin{equation}
	g_{C_n}(\theta)
	=
	S_{C_n}
	\Big(
	AJ^{(j)}
	(\theta-\theta^{(j)})
	-
	d^{(j)}
	\Big).
	\label{eq:A5}
\end{equation}
Then $\mathcal{L}_{C_n}(\theta)\approx \|g_{C_n}(\theta)\|_2$. On a local neighborhood where $g_{C_n}(\theta)\neq 0$, the gradient is
\begin{equation}
	\nabla \mathcal{L}_{C_n}(\theta)
	=
	\frac{
		(J^{(j)})^H A^H S_{C_n}^H g_{C_n}(\theta)
	}{
		\|g_{C_n}(\theta)\|_2
	}.
	\label{eq:A6}
\end{equation}
Assume that the local residual is bounded away from zero,
\begin{equation}
	\|g_{C_n}(\theta)\|_2
	\ge
	\delta_{C_n}
	>
	0.
	\label{eq:A7}
\end{equation}
Then $\nabla \mathcal{L}_{C_n}$ is locally Lipschitz continuous. Let $\beta_{C_n}>0$ denote a local Lipschitz constant. We further assume that $\mathcal{L}_{C_n}$ satisfies the PL inequality~\cite{xiao2023generalized, nejma2025polyak}
\begin{equation}
	\frac{1}{2}
	\left\|
	\nabla \mathcal{L}_{C_n}(\theta)
	\right\|_2^2
	\ge
	\mu_{C_n}
	\big(
	\mathcal{L}_{C_n}(\theta)-\mathcal{L}_{C_n}^{*}
	\big),
	\label{eq:A8}
\end{equation}
where $\mathcal{L}_{C_n}^{*}=\inf_{\theta}\mathcal{L}_{C_n}(\theta)$, and $\mu_{C_n}>0$ is a local effective PL curvature. The constants $\beta_{C_n}$ and $\mu_{C_n}$ are local trajectory-dependent quantities, not global Hessian eigenvalues of a quadratic objective. In the early stages of CogGen, the scheduler tends to expose measurements that are structurally informative and readily fitable by the current generator, while deferring measurements that are harder to fit or more sensitive to noise. Under this early-stage fitability preference, the weighted local surrogate may admit a more favorable local geometry than uniform fitting along the identifiable trajectory.

\noindent\textbf{Assumption A.1. Local early-stage fitting condition.}
Let $n_0$ be the last stage index of the early stage phase. Define
\begin{equation}
	\mu_{\mathrm{early}}
	=
	\inf_{1\le n\le n_0}
	\mu_{C_n}.
	\label{eq:A9}
\end{equation}
Let $\mu_{\mathrm{uniform}}$ denote the corresponding local effective PL curvature under uniform fitting, i.e., $S_{C_n}\equiv I$. We assume that, along the locally identifiable optimization trajectory,
\begin{equation}
	\mu_{\mathrm{early}}
	>
	\mu_{\mathrm{uniform}}.
	\label{eq:A10}
\end{equation}
This assumption does not state that down-weighting measurements universally improves conditioning. It only describes the early-stage regime where selecting readily fitable measurements makes the local surrogate better aligned with the current generator representation.

Consider the stage-wise gradient descent update
\begin{equation}
	\theta^{(\ell+1)}
	=
	\theta^{(\ell)}
	-
	\tau
	\nabla \mathcal{L}_{C_n}
	\big(\theta^{(\ell)}\big),
	\qquad
	0<\tau\le \frac{1}{\beta_{C_n}}.
	\label{eq:A11}
\end{equation}

\noindent\textbf{Lemma A.1. Stage-wise local linear convergence.}
Assume that the iterates remain in the local neighborhood where \eqref{eq:A7} and \eqref{eq:A8} hold. Under \eqref{eq:A8} and \eqref{eq:A11}, with $0<\tau\mu_{C_n}\le1$, the iterates satisfy
\begin{equation}
	\begin{aligned}
		&
		\mathcal{L}_{C_n}
		\big(\theta^{(\ell)}\big)
		-
		\mathcal{L}_{C_n}^{*}
		\\
		&\le
		\big(1-\tau\mu_{C_n}\big)^{\ell}
		\Big[
		\mathcal{L}_{C_n}
		\big(\theta^{(0)}\big)
		-
		\mathcal{L}_{C_n}^{*}
		\Big].
	\end{aligned}
	\label{eq:A12}
\end{equation}

\noindent\textit{Proof.}
By local $\beta_{C_n}$-smoothness,
\begin{equation}
	\begin{aligned}
		&
		\mathcal{L}_{C_n}
		\big(\theta^{(\ell+1)}\big)
		\\
		&\le
		\mathcal{L}_{C_n}
		\big(\theta^{(\ell)}\big)
		-
		\tau
		\left(
		1-\frac{\tau\beta_{C_n}}{2}
		\right)
		\left\|
		\nabla\mathcal{L}_{C_n}
		\big(\theta^{(\ell)}\big)
		\right\|_2^2.
	\end{aligned}
	\label{eq:A13}
\end{equation}
Since $0<\tau\le 1/\beta_{C_n}$,
\begin{equation}
	1-
	\frac{\tau\beta_{C_n}}{2}
	\ge
	\frac{1}{2}.
	\label{eq:A14}
\end{equation}
Combining \eqref{eq:A13}, \eqref{eq:A14}, and the PL inequality \eqref{eq:A8} gives
\begin{equation}
	\begin{aligned}
		&
		\mathcal{L}_{C_n}
		\big(\theta^{(\ell+1)}\big)
		-
		\mathcal{L}_{C_n}^{*}
		\\
		&\le
		\big(1-\tau\mu_{C_n}\big)
		\Big[
		\mathcal{L}_{C_n}
		\big(\theta^{(\ell)}\big)
		-
		\mathcal{L}_{C_n}^{*}
		\Big].
	\end{aligned}
	\label{eq:A15}
\end{equation}
Iterating \eqref{eq:A15} proves \eqref{eq:A12}. This completes the proof. $\square$

\noindent\textbf{Theorem A.1. Reduced local sufficient-iteration bound.}
Fix any relative accuracy level $\xi\in(0,1)$. Suppose that the assumptions of Lemma A.1 hold for both the early stage-wise weighted surrogates and the corresponding uniform-fitting surrogate, with the same admissible step size $\tau$. Under Assumption A.1, a sufficient number of gradient steps for CogGen at any early stage $n\le n_0$ is
\begin{equation}
	\ell_{\mathrm{CogGen}}^{\mathrm{suf}}
	=
	\left\lceil
	\frac{
		\log(1/\xi)
	}{
		\tau\mu_{\mathrm{early}}
	}
	\right\rceil.
	\label{eq:A16}
\end{equation}
For uniform fitting, the corresponding sufficient number of steps is
\begin{equation}
	\ell_{\mathrm{uniform}}^{\mathrm{suf}}
	=
	\left\lceil
	\frac{
		\log(1/\xi)
	}{
		\tau\mu_{\mathrm{uniform}}
	}
	\right\rceil.
	\label{eq:A17}
\end{equation}
Before integer rounding, the CogGen sufficient-iteration bound is strictly smaller than the uniform-fitting bound. After integer rounding,
\begin{equation}
	\ell_{\mathrm{CogGen}}^{\mathrm{suf}}
	\le
	\ell_{\mathrm{uniform}}^{\mathrm{suf}}.
	\label{eq:A18}
\end{equation}

\noindent\textit{Proof.}
If
\begin{equation}
	\mathcal{L}_{C_n}(\theta^{(0)})
	-
	\mathcal{L}_{C_n}^{*}
	=
	0,
	\label{eq:A19}
\end{equation}
then the desired relative accuracy has already been reached. Otherwise, from \eqref{eq:A12},
\begin{equation}
	\frac{
		\mathcal{L}_{C_n}(\theta^{(\ell)})
		-
		\mathcal{L}_{C_n}^{*}
	}{
		\mathcal{L}_{C_n}(\theta^{(0)})
		-
		\mathcal{L}_{C_n}^{*}
	}
	\le
	\big(1-\tau\mu_{C_n}\big)^{\ell}.
	\label{eq:A20}
\end{equation}
Using $1-x\le e^{-x}$,
\begin{equation}
	\big(1-\tau\mu_{C_n}\big)^{\ell}
	\le
	\exp
	\big(-\tau\mu_{C_n}\ell\big).
	\label{eq:A21}
\end{equation}
Thus, to make the relative objective gap no larger than $\rho$, it suffices to require
\begin{equation}
	\ell
	\ge
	\frac{
		\log(1/\xi)
	}{
		\tau\mu_{C_n}
	}.
	\label{eq:A22}
\end{equation}
For all early stages $n\le n_0$, $\mu_{C_n}\ge\mu_{\mathrm{early}}$. Hence \eqref{eq:A16} is sufficient for CogGen at any early stage. For uniform fitting, applying the same argument with $\mu_{\mathrm{uniform}}$ gives \eqref{eq:A17}. Since $\mu_{\mathrm{early}}>\mu_{\mathrm{uniform}}$,
\begin{equation}
	\frac{
		\log(1/\xi)
	}{
		\tau\mu_{\mathrm{early}}
	}
	<
	\frac{
		\log(1/\xi)
	}{
		\tau\mu_{\mathrm{uniform}}
	}.
	\label{eq:A23}
\end{equation}
Therefore, the unrounded sufficient-iteration bound of CogGen is strictly smaller than that of uniform fitting. Since the ceiling function is monotone nondecreasing,
\begin{equation}
	\ell_{\mathrm{CogGen}}^{\mathrm{suf}}
	\le
	\ell_{\mathrm{uniform}}^{\mathrm{suf}}.
	\label{eq:A24}
\end{equation}
This proves the stated local sufficient-iteration comparison. $\square$

\subsection{Cumulative Noise-Amplification Bound of CogGen}

We next analyze how stage-wise measurement weighting affects the accumulation of measurement noise during iterative reconstruction. In conventional FU-DGMs, prolonged fitting of noisy and ill-posed measurements may imprint measurement noise into the reconstructed image. The following analysis shows that CogGen can reduce the effective noise component injected during early updates, leading to a smaller upper bound on cumulative noise amplification under local stability conditions. Consider the stage-wise image-domain surrogate
\begin{equation}
	L_{C_n}(x)
	=
	\frac{
		\left\|
		S_{C_n}(Ax-y)
		\right\|_2
	}{
		\left\|
		S_{C_n}y
		\right\|_2
	}.
	\label{eq:B1}
\end{equation}
Let
\begin{equation}
	W_{C_n}
	=
	S_{C_n}^{H}S_{C_n}.
	\label{eq:B2}
\end{equation}
On a local neighborhood where $S_{C_n}(Ax-y)\neq0$, the gradient direction is proportional to
\begin{equation}
	A^{H}W_{C_n}(Ax-y).
	\label{eq:B3}
\end{equation}

To expose the noise-propagation mechanism, we consider the following rescaled Landweber-type surrogate dynamics:
\begin{equation}
	x_{C_n}^{(\ell+1)}
	=
	x_{C_n}^{(\ell)}
	-
	\alpha_{C_n}^{(\ell)}
	A^{H}W_{C_n}
	\big(Ax_{C_n}^{(\ell)}-y\big).
	\label{eq:B4}
\end{equation}
The rescaling factor induced by the non-squared normalized fidelity is
\begin{equation}
	\alpha_{C_n}^{(\ell)}
	=
	\frac{
		\tau
	}{
		\left\|
		S_{C_n}
		\big(Ax_{C_n}^{(\ell)}-y\big)
		\right\|_2
		\left\|
		S_{C_n}y
		\right\|_2
	}.
	\label{eq:B5}
\end{equation}
This surrogate is used to describe how the stage-wise weighting operator modulates the noise injected through the data-consistency update.

Let
\begin{equation}
	e_{C_n}^{(\ell)}
	=
	x_{C_n}^{(\ell)}
	-
	x^{*}.
	\label{eq:B6}
\end{equation}
Using the noisy observation model
\begin{equation}
	y
	=
	Ax^{*}
	+
	\varepsilon,
	\label{eq:B7}
\end{equation}
we obtain the error recursion
\begin{equation}
	\begin{aligned}
		e_{C_n}^{(\ell+1)}
		=
		&
		\left(
		I
		-
		\alpha_{C_n}^{(\ell)}
		A^{H}W_{C_n}A
		\right)
		e_{C_n}^{(\ell)}
		\\
		&+
		\alpha_{C_n}^{(\ell)}
		A^{H}W_{C_n}\varepsilon .
	\end{aligned}
	\label{eq:B8}
\end{equation}
Define
\begin{equation}
	M_{C_n}^{(\ell)}
	=
	I
	-
	\alpha_{C_n}^{(\ell)}
	A^{H}W_{C_n}A.
	\label{eq:B9}
\end{equation}

Since $A$ is undersampled, strict contraction is not expected on the entire image space. We therefore restrict the analysis to the locally identifiable subspace $\mathcal{X}_{C_n}$. Along this subspace, assume that
\begin{equation}
	0
	<
	\alpha_{C_n}^{(\ell)}
	\le
	\alpha_{\max},
	\label{eq:B10}
\end{equation}
and
\begin{equation}
	\left\|
	M_{C_n}^{(\ell)}
	\big|_{\mathcal{X}_{C_n}}
	\right\|_2
	\le
	\rho_{C_n},
	\qquad
	0<\rho_{C_n}<1.
	\label{eq:B11}
\end{equation}

Let $L(n)$ denote the number of inner iterations at stage $C_n$.

\noindent\textbf{Lemma A.2. Local cumulative noise-propagation bound.}
Under the error recursion \eqref{eq:B8} and the local stability conditions \eqref{eq:B10}--\eqref{eq:B11}, the noise-induced component of the final reconstruction error satisfies
\begin{equation}
	\left\|
	e_{\mathrm{noise}}^{(T)}
	\right\|_2
	\le
	\alpha_{\max}
	\|A\|_2
	\sum_{n=1}^{T}
	\Gamma_n
	\left\|
	W_{C_n}\varepsilon
	\right\|_2,
	\label{eq:B12}
\end{equation}
where
\begin{equation}
	\Gamma_n
	=
	\left(
	\prod_{k=n+1}^{T}
	\rho_{C_k}^{L(k)}
	\right)
	\left(
	\sum_{\ell=0}^{L(n)-1}
	\rho_{C_n}^{\ell}
	\right).
	\label{eq:B13}
\end{equation}

\noindent\textit{Proof.}
Let $e_{\mathrm{noise}}$ denote the component of the error generated by the measurement noise term in \eqref{eq:B8}. Taking norms on the locally identifiable subspace and using \eqref{eq:B10}--\eqref{eq:B11} gives, within stage $C_n$,
\begin{equation}
	\left\|
	e_{\mathrm{noise},C_n}^{(\ell+1)}
	\right\|_2
	\le
	\rho_{C_n}
	\left\|
	e_{\mathrm{noise},C_n}^{(\ell)}
	\right\|_2
	+
	\alpha_{\max}
	\|A\|_2
	\left\|
	W_{C_n}\varepsilon
	\right\|_2 .
	\label{eq:B14}
\end{equation}
Thus, the cumulative noise injected during the $L(n)$ inner iterations of stage $C_n$ is bounded, at the exit of this stage, by
\begin{equation}
	\alpha_{\max}
	\|A\|_2
	\left(
	\sum_{\ell=0}^{L(n)-1}
	\rho_{C_n}^{\ell}
	\right)
	\left\|
	W_{C_n}\varepsilon
	\right\|_2 .
	\label{eq:B15}
\end{equation}
This contribution is further propagated through later stages $C_{n+1},\ldots,C_T$, which gives the multiplicative factor $\prod_{k=n+1}^{T}\rho_{C_k}^{L(k)}$. Summing the propagated contributions over all stages yields \eqref{eq:B12} with $\Gamma_n$ defined in \eqref{eq:B13}. This completes the proof. $\square$

For a uniform-fitting FU-DGM baseline, all acquired measurements participate in the data-consistency term throughout the inference process. For notational simplicity, this representative uniform-fitting baseline is denoted as
\begin{equation}
	W_{C_n}^{\mathrm{uniform}}
	\equiv
	I_M,
	\qquad
	n=1,\ldots,T.
	\label{eq:B17}
\end{equation}
Thus,
\begin{equation}
	\left\|
	W_{C_n}^{\mathrm{uniform}}
	\varepsilon
	\right\|_2
	=
	\|\varepsilon\|_2.
	\label{eq:B18}
\end{equation}

\noindent\textbf{Assumption A.2. Early-stage noise attenuation.}
For CogGen, the early stages defer a subset of difficult or noise-sensitive measurements. Assume that there exist $n_0<T$ and $q\in(0,1)$ such that
\begin{equation}
	\left\|
	W_{C_n}\varepsilon
	\right\|_2
	\le
	q\|\varepsilon\|_2,
	\qquad
	n=1,\ldots,n_0,
	\label{eq:B19}
\end{equation}
and
\begin{equation}
	\left\|
	W_{C_n}\varepsilon
	\right\|_2
	\le
	\|\varepsilon\|_2,
	\qquad
	n=n_0+1,\ldots,T.
	\label{eq:B20}
\end{equation}
This captures the intended effect of progressive exposure: early updates avoid fitting all noisy measurements at once, while later updates gradually incorporate more complete data information.

\noindent\textbf{Theorem A.2. Smaller cumulative noise-amplification bound.}
Assume that Lemma A.2 holds. To isolate the noise-injection effect, suppose that CogGen and the representative uniform-fitting baseline are compared under the same stage lengths and comparable local stability factors summarized by $\Gamma_n$. Under Assumption A.2, the CogGen noise-amplification bound is
\begin{equation}
	\begin{aligned}
		\mathcal{B}_{\mathrm{CogGen}}(T)
		=
		&
		\alpha_{\max}
		\|A\|_2
		\|\varepsilon\|_2
		\\
		&\times
		\left(
		q\sum_{n=1}^{n_0}\Gamma_n
		+
		\sum_{n=n_0+1}^{T}\Gamma_n
		\right),
	\end{aligned}
	\label{eq:B21}
\end{equation}
whereas the corresponding uniform-fitting bound is
\begin{equation}
	\begin{aligned}
		\mathcal{B}_{\mathrm{uniform}}(T)
		=
		&
		\alpha_{\max}
		\|A\|_2
		\|\varepsilon\|_2
		\\
		&\times
		\left(
		\sum_{n=1}^{n_0}\Gamma_n
		+
		\sum_{n=n_0+1}^{T}\Gamma_n
		\right).
	\end{aligned}
	\label{eq:B22}
\end{equation}
If
\begin{equation}
	\sum_{n=1}^{n_0}\Gamma_n
	>
	0,
	\label{eq:B23}
\end{equation}
then
\begin{equation}
	\mathcal{B}_{\mathrm{CogGen}}(T)
	<
	\mathcal{B}_{\mathrm{uniform}}(T).
	\label{eq:B24}
\end{equation}
Consequently,
\begin{equation}
	\left\|
	e_{\mathrm{noise,CogGen}}^{(T)}
	\right\|_2
	\le
	\mathcal{B}_{\mathrm{CogGen}}(T)
	<
	\mathcal{B}_{\mathrm{uniform}}(T).
	\label{eq:B25}
\end{equation}

\noindent\textit{Proof.}
By Lemma A.2 and Assumption A.2,
\begin{equation}
	\begin{aligned}
		\left\|
		e_{\mathrm{noise,CogGen}}^{(T)}
		\right\|_2
		\le
		&
		\alpha_{\max}
		\|A\|_2
		\|\varepsilon\|_2
		\\
		&\times
		\left(
		q\sum_{n=1}^{n_0}\Gamma_n
		+
		\sum_{n=n_0+1}^{T}\Gamma_n
		\right).
	\end{aligned}
	\label{eq:B26}
\end{equation}
This gives \eqref{eq:B21}. For uniform fitting, using \eqref{eq:B18} in Lemma A.2 gives
\begin{equation}
	\begin{aligned}
		\left\|
		e_{\mathrm{noise,uniform}}^{(T)}
		\right\|_2
		\le
		&
		\alpha_{\max}
		\|A\|_2
		\|\varepsilon\|_2
		\\
		&\times
		\left(
		\sum_{n=1}^{n_0}\Gamma_n
		+
		\sum_{n=n_0+1}^{T}\Gamma_n
		\right),
	\end{aligned}
	\label{eq:B27}
\end{equation}
which gives \eqref{eq:B22}. Since $q<1$ and \eqref{eq:B23} holds,
\begin{equation}
	\begin{aligned}
		&
		\mathcal{B}_{\mathrm{uniform}}(T)
		-
		\mathcal{B}_{\mathrm{CogGen}}(T)
		\\
		&=
		\alpha_{\max}
		\|A\|_2
		\|\varepsilon\|_2
		(1-q)
		\sum_{n=1}^{n_0}\Gamma_n
		>
		0.
	\end{aligned}
	\label{eq:B28}
\end{equation}
Therefore, $\mathcal{B}_{\mathrm{CogGen}}(T)<\mathcal{B}_{\mathrm{uniform}}(T)$, and \eqref{eq:B25} follows from \eqref{eq:B26}. $\square$

Therefore, when early-stage weighting reduces the effective noise component and the local stability factors are comparable, CogGen admits a smaller upper bound on the cumulative noise-amplification term than the corresponding uniform fitting. This explains why CogGen can suppress noise imprinting and preserve image details more effectively during finite-budget reconstruction.
\balance
\bibliographystyle{IEEEtran}
\bibliography{main}

@String(ICASSP=	{ICASSP})

@article{yaman2020self,
	title={Self-supervised learning of physics-guided reconstruction neural networks without fully sampled reference data},
	author={Yaman, Burhaneddin and Hosseini, Seyed Amir Hossein and Moeller, Steen and Ellermann, Jutta and U{\u{g}}urbil, K{\^a}mil and Ak{\c{c}}akaya, Mehmet},
	journal={Magnetic resonance in medicine},
	volume={84},
	number={6},
	pages={3172--3191},
	year={2020},
	publisher={Wiley Online Library}
}

@article{shu2023cmw,
	title={Cmw-net: Learning a class-aware sample weighting mapping for robust deep learning},
	author={Shu, Jun and Yuan, Xiang and Meng, Deyu and Xu, Zongben},
	journal={IEEE Transactions on Pattern Analysis and Machine Intelligence},
	volume={45},
	number={10},
	pages={11521--11539},
	year={2023},
	publisher={IEEE}
}

@article{de2024role,
	title={The role of MRI in radiotherapy planning: a narrative review “from head to toe”},
	author={De Pietro, Simona and Di Martino, Giulia and Caroprese, Mara and Barillaro, Angela and Cocozza, Sirio and Pacelli, Roberto and Cuocolo, Renato and Ugga, Lorenzo and Briganti, Francesco and Brunetti, Arturo and others},
	journal={Insights into Imaging},
	volume={15},
	number={1},
	pages={255},
	year={2024},
	publisher={Springer}
}

@article{buskulic2024convergence,
	title={Convergence and recovery guarantees of unsupervised neural networks for inverse problems},
	author={Buskulic, Nathan and Fadili, Jalal and Qu{\'e}au, Yvain},
	journal={Journal of Mathematical Imaging and Vision},
	volume={66},
	number={4},
	pages={584--605},
	year={2024},
	publisher={Springer}
}

@article{chen2023cognitive,
	title={A cognitive load theory approach to defining and measuring task complexity through element interactivity},
	author={Chen, Ouhao and Paas, Fred and Sweller, John},
	journal={Educational Psychology Review},
	volume={35},
	number={2},
	pages={63},
	year={2023},
	publisher={Springer}
}

@article{zhou2023investigating,
	title={Investigating the sample weighting mechanism using an interpretable weighting framework},
	author={Zhou, Xiaoling and Wu, Ou and Li, Mengyang},
	journal={IEEE Transactions on Knowledge and Data Engineering},
	volume={36},
	number={5},
	pages={2041--2055},
	year={2023},
	publisher={IEEE}
}

@article{wu2025data,
	title={Data optimization in deep learning: A survey},
	author={Wu, Ou and Yao, Rujing},
	journal={IEEE Transactions on Knowledge and Data Engineering},
	volume={37},
	number={5},
	pages={2356--2375},
	year={2025},
	publisher={IEEE}
}

@inproceedings{liu2025uhd,
	title={Uhd-processer: Unified uhd image restoration with progressive frequency learning and degradation-aware prompts},
	author={Liu, Yidi and Li, Dong and Fu, Xueyang and Lu, Xin and Huang, Jie and Zha, Zheng-Jun},
	booktitle={Proceedings of the Computer Vision and Pattern Recognition Conference},
	pages={23121--23130},
	year={2025}
}

@article{benfenati2025early,
	title={Early stopping strategies in Deep Image Prior},
	author={Benfenati, Alessandro and Catozzi, Ambra and Franchini, Giorgia and Porta, Federica},
	journal={Soft Computing},
	volume={29},
	number={8},
	pages={4153--4174},
	year={2025},
	publisher={Springer}
}

@inproceedings{molaei2023implicit,
	title={Implicit neural representation in medical imaging: A comparative survey},
	author={Molaei, Amirali and Aminimehr, Amirhossein and Tavakoli, Armin and Kazerouni, Amirhossein and Azad, Bobby and Azad, Reza and Merhof, Dorit},
	booktitle={Proceedings of the IEEE/CVF International Conference on Computer Vision},
	pages={2381--2391},
	year={2023}
}

@article{xiao2023generalized,
	title={A generalized alternating method for bilevel learning under the {Polyak--{\L}ojasiewicz} condition},
	author={Xiao, Quan and Lu, Songtao and Chen, Tianyi},
	journal={arXiv preprint arXiv:2306.02422},
	year={2023}
}

@article{nejma2025polyak,
	title={{Polyak--{\L}ojasiewicz} inequality is essentially no more general than strong convexity for $ C^{2}$ functions},
	author={Nejma, Aziz Ben},
	journal={arXiv preprint arXiv:2512.05285},
	year={2025}
}

@article{golikov2022neural,
	title={Neural tangent kernel: A survey},
	author={Golikov, Eugene and Pokonechnyy, Eduard and Korviakov, Vladimir},
	journal={arXiv preprint arXiv:2208.13614},
	year={2022}
}

@article{ref41,
  title={MoDL: Model-based deep learning architecture for inverse problems},
  author={Aggarwal, Hemant K and Mani, Merry P and Jacob, Mathews},
  journal={IEEE transactions on medical imaging},
  volume={38},
  number={2},
  pages={394--405},
  year={2018},
  publisher={IEEE}
}

@article{shi2023greedy,
	title={Greedy guarantees for minimum submodular cost submodular/non-submodular cover problem},
	author={Shi, Majun and Yang, Zishen and Wang, Wei},
	journal={Journal of Combinatorial Optimization},
	volume={45},
	number={1},
	pages={8},
	year={2023},
	publisher={Springer}
}

@article{shi2025weak,
	title={Weak submodularity implies localizability: Local search for constrained non-submodular function maximization},
	author={Shi, Majun and Zhu, Qingyong and Liu, Bei and Li, Yuchao},
	journal={Discrete Mathematics},
	volume={348},
	number={2},
	pages={114287},
	year={2025},
	publisher={Elsevier}
}

@article{el2022data,
	title={Data-efficient structured pruning via submodular optimization},
	author={El Halabi, Marwa and Srinivas, Suraj and Lacoste-Julien, Simon},
	journal={Advances in Neural Information Processing Systems},
	volume={35},
	pages={36613--36626},
	year={2022}
}

@article{yang2021learning,
	title={Learning interpretable decision rule sets: A submodular optimization approach},
	author={Yang, Fan and He, Kai and Yang, Linxiao and Du, Hongxia and Yang, Jingbang and Yang, Bo and Sun, Liang},
	journal={Advances in Neural Information Processing Systems},
	volume={34},
	pages={27890--27902},
	year={2021}
}

@article{hussain2022modern,
	title={Modern diagnostic imaging technique applications and risk factors in the medical field: a review},
	author={Hussain, Shah and Mubeen, Iqra and Ullah, Niamat and Shah, Syed Shahab Ud Din and Khan, Bakhtawar Abduljalil and Zahoor, Muhammad and Ullah, Riaz and Khan, Farhat Ali and Sultan, Mujeeb A},
	journal={BioMed research international},
	volume={2022},
	number={1},
	pages={5164970},
	year={2022},
	publisher={Wiley Online Library}
}

@article{sandino2020compressed,
	title={Compressed sensing: From research to clinical practice with deep neural networks: Shortening scan times for magnetic resonance imaging},
	author={Sandino, Christopher M and Cheng, Joseph Y and Chen, Feiyu and Mardani, Morteza and Pauly, John M and Vasanawala, Shreyas S},
	journal={IEEE signal processing magazine},
	volume={37},
	number={1},
	pages={117--127},
	year={2020},
	publisher={IEEE}
}

@article{sweller2023development,
	title={The development of cognitive load theory: Replication crises and incorporation of other theories can lead to theory expansion},
	author={Sweller, John},
	journal={Educational Psychology Review},
	volume={35},
	number={4},
	pages={95},
	year={2023},
	publisher={Springer}
}

@article{ref45,
  title={Image reconstruction via autoencoding sequential deep image prior},
  author={Alkhouri, Ismail and Liang, Shijun and Bell, Evan and Qu, Qing and Wang, Rongrong and Ravishankar, Saiprasad},
  journal={Advances in Neural Information Processing Systems},
  volume={37},
  pages={18988--19012},
  year={2024}
}

@article{yoo2021time,
	title={Time-dependent deep image prior for dynamic MRI},
	author={Yoo, Jaejun and Jin, Kyong Hwan and Gupta, Harshit and Yerly, Jerome and Stuber, Matthias and Unser, Michael},
	journal={IEEE Transactions on Medical Imaging},
	volume={40},
	number={12},
	pages={3337--3348},
	year={2021},
	publisher={IEEE}
}

@ARTICLE{10374196,
	author={Zhu, Qingyong and Liu, Bei and Cui, Zhuo-Xu and Cao, Chentao and Yan, Xiaomeng and Liu, Yuanyuan and Cheng, Jing and Zhou, Yihang and Zhu, Yanjie and Wang, Haifeng and Zeng, Hongwu and Liang, Dong},
	journal={IEEE Journal of Biomedical and Health Informatics}, 
	title={PEARL: Cascaded Self-Supervised Cross-Fusion Learning for Parallel MRI Acceleration}, 
	year={2025},
	volume={29},
	number={5},
	pages={3086-3097},
	doi={10.1109/JBHI.2023.3347355}}

@article{darestani2021accelerated,
	title={Accelerated MRI with un-trained neural networks},
	author={Darestani, Mohammad Zalbagi and Heckel, Reinhard},
	journal={IEEE Transactions on Computational Imaging},
	volume={7},
	pages={724--733},
	year={2021},
	publisher={IEEE}
}

@ARTICLE{9173689,
	author={Oh, Gyutaek and Sim, Byeongsu and Chung, HyungJin and Sunwoo, Leonard and Ye, Jong Chul},
	journal={IEEE Transactions on Computational Imaging}, 
	title={Unpaired Deep Learning for Accelerated MRI Using Optimal Transport Driven CycleGAN}, 
	year={2020},
	volume={6},
	number={},
	pages={1285-1296},
	doi={10.1109/TCI.2020.3018562}}

@ARTICLE{8327637,
	author={Quan, Tran Minh and Nguyen-Duc, Thanh and Jeong, Won-Ki},
	journal={IEEE Transactions on Medical Imaging}, 
	title={Compressed Sensing MRI Reconstruction Using a Generative Adversarial Network With a Cyclic Loss}, 
	year={2018},
	volume={37},
	number={6},
	pages={1488-1497},
	doi={10.1109/TMI.2018.2820120}}

@ARTICLE{8233175,
	author={Yang, Guang and Yu, Simiao and Dong, Hao and Slabaugh, Greg and Dragotti, Pier Luigi and Ye, Xujiong and Liu, Fangde and Arridge, Simon and Keegan, Jennifer and Guo, Yike and Firmin, David},
	journal={IEEE Transactions on Medical Imaging}, 
	title={DAGAN: Deep De-Aliasing Generative Adversarial Networks for Fast Compressed Sensing MRI Reconstruction}, 
	year={2018},
	volume={37},
	number={6},
	pages={1310-1321},
	doi={10.1109/TMI.2017.2785879}}

@article{li2021high,
	title={High quality and fast compressed sensing MRI reconstruction via edge-enhanced dual discriminator generative adversarial network},
	author={Li, Yixuan and Li, Jie and Ma, Fengfei and Du, Shuangli and Liu, Yiguang},
	journal={Magnetic Resonance Imaging},
	volume={77},
	pages={124--136},
	year={2021},
	publisher={Elsevier}
}

@article{ruthotto2021introduction,
	title={An introduction to deep generative modeling},
	author={Ruthotto, Lars and Haber, Eldad},
	journal={GAMM-Mitteilungen},
	volume={44},
	number={2},
	pages={e202100008},
	year={2021},
	publisher={Wiley Online Library}
}

@article{suzuki2022survey,
	title={A survey of multimodal deep generative models},
	author={Suzuki, Masahiro and Matsuo, Yutaka},
	journal={Advanced Robotics},
	volume={36},
	number={5-6},
	pages={261--278},
	year={2022},
	publisher={Taylor \& Francis}
}

@inproceedings{ref48,
  title={Curriculum learning},
  author={Bengio, Yoshua and Louradour, J{\'e}r{\^o}me and Collobert, Ronan and Weston, Jason},
  booktitle={Proceedings of the 26th annual international conference on machine learning},
  pages={41--48},
  year={2009}
}

@article{ref46,
  title={Decoupled algorithm for MRI reconstruction using nonlocal block matching model: BM3D-MRI},
  author={Eksioglu, Ender M},
  journal={Journal of Mathematical Imaging and Vision},
  volume={56},
  number={3},
  pages={430--440},
  year={2016},
  publisher={Springer}
}

@article{ref42,
  title={fastMRI: A publicly available raw k-space and DICOM dataset of knee images for accelerated MR image reconstruction using machine learning},
  author={Knoll, Florian and Zbontar, Jure and Sriram, Anuroop and Muckley, Matthew J and Bruno, Mary and Defazio, Aaron and Parente, Marc and Geras, Krzysztof J and Katsnelson, Joe and Chandarana, Hersh and others},
  journal={Radiology: Artificial Intelligence},
  volume={2},
  number={1},
  pages={e190007},
  year={2020},
  publisher={Radiological Society of North America}
}

@inproceedings{wang2024progressive,
	title={Progressive divide-and-conquer via subsampling decomposition for accelerated mri},
	author={Wang, Chong and Guo, Lanqing and Wang, Yufei and Cheng, Hao and Yu, Yi and Wen, Bihan},
	booktitle={Proceedings of the IEEE/CVF Conference on Computer Vision and Pattern Recognition},
	pages={25128--25137},
	year={2024}
}

@inproceedings{ref28,
  title={Self-paced learning for latent variable models},
  author={Kumar, M and Packer, Benjamin and Koller, Daphne},
  journal={Advances in neural information processing systems},
  volume={23},
  year={2010}
}

@article{zhang2019leveraging,
	title={Leveraging prior-knowledge for weakly supervised object detection under a collaborative self-paced curriculum learning framework},
	author={Zhang, Dingwen and Han, Junwei and Zhao, Long and Meng, Deyu},
	journal={International Journal of Computer Vision},
	volume={127},
	number={4},
	pages={363--380},
	year={2019},
	publisher={Springer}
}

@article{kim2018screenernet,
	title={Screenernet: Learning self-paced curriculum for deep neural networks},
	author={Kim, Tae-Hoon and Choi, Jonghyun},
	journal={arXiv preprint arXiv:1801.00904},
	year={2018}
}

@article{choi2025brain,
	title={Brain-Guided Self-Paced Curriculum Learning for Adaptive Human--Machine Interfaces},
	author={Choi, Yeon-Woo and Shin, Hye-Bin and Lee, Seong-Whan},
	journal={IEEE Transactions on Systems, Man, and Cybernetics: Systems},
	year={2025},
	publisher={IEEE}
}

@article{saragadam2024deeptensor,
	title={DeepTensor: Low-rank tensor decomposition with deep network priors},
	author={Saragadam, Vishwanath and Balestriero, Randall and Veeraraghavan, Ashok and Baraniuk, Richard G},
	journal={IEEE Transactions on Pattern Analysis and Machine Intelligence},
	year={2024},
	publisher={IEEE}
}

@inproceedings{ref43,
  title={Image restoration using total variation regularized deep image prior},
  author={Liu, Jiaming and Sun, Yu and Xu, Xiaojian and Kamilov, Ulugbek S},
  booktitle={ICASSP 2019-2019 IEEE International Conference on Acoustics, Speech and Signal Processing (ICASSP)},
  pages={7715--7719},
  year={2019},
  organization={Ieee}
}

@article{ref47,
  title={An unsupervised method for MRI recovery: deep image prior with structured sparsity},
  author={Sultan, Muhammad Ahmad and Chen, Chong and Liu, Yingmin and Gil, Katarzyna and Zareba, Karolina and Ahmad, Rizwan},
  journal={Magnetic Resonance Materials in Physics, Biology and Medicine},
  pages={1--13},
  year={2025},
  publisher={Springer}
}

@inproceedings{ref26,
  title={Deep image prior},
  author={Ulyanov, Dmitry and Vedaldi, Andrea and Lempitsky, Victor},
  booktitle={Proceedings of the IEEE conference on computer vision and pattern recognition},
  pages={9446--9454},
  year={2018}
}

@article{eksioglu2016decoupled,
	title={Decoupled algorithm for MRI reconstruction using nonlocal block matching model: BM3D-MRI},
	author={Eksioglu, Ender M},
	journal={Journal of Mathematical Imaging and Vision},
	volume={56},
	pages={430--440},
	year={2016},
	publisher={Springer}
}

@article{cohen2021regularization,
	title={Regularization by denoising via fixed-point projection (RED-PRO)},
	author={Cohen, Regev and Elad, Michael and Milanfar, Peyman},
	journal={SIAM Journal on Imaging Sciences},
	volume={14},
	number={3},
	pages={1374--1406},
	year={2021},
	publisher={SIAM}
}

@article{wang2021early,
	title={Early stopping for deep image prior},
	author={Wang, Hengkang and Li, Taihui and Zhuang, Zhong and Chen, Tiancong and Liang, Hengyue and Sun, Ju},
	journal={arXiv preprint arXiv:2112.06074},
	year={2021}
}

@article{shen2022nerp,
	title={NeRP: implicit neural representation learning with prior embedding for sparsely sampled image reconstruction},
	author={Shen, Liyue and Pauly, John and Xing, Lei},
	journal={IEEE Transactions on Neural Networks and Learning Systems},
	volume={35},
	number={1},
	pages={770--782},
	year={2022},
	publisher={IEEE}
}

@article{heckel2018deep,
	title={Deep decoder: Concise image representations from untrained non-convolutional networks},
	author={Heckel, Reinhard and Hand, Paul},
	journal={arXiv preprint arXiv:1810.03982},
	year={2018},
}

@inproceedings{jagatap2019phase,
	title={Phase retrieval using untrained neural network priors},
	author={Jagatap, Gauri and Hegde, Chinmay},
	booktitle={NeurIPS 2019 Workshop on Solving Inverse Problems with Deep Networks},
	year={2019}
}

@inproceedings{buskulic2023convergence,
	title={Convergence guarantees of overparametrized wide deep inverse prior},
	author={Buskulic, Nathan and Qu{\'e}au, Yvain and Fadili, Jalal},
	booktitle={International Conference on Scale Space and Variational Methods in Computer Vision},
	pages={406--417},
	year={2023},
	organization={Springer}
}

@misc{heckel2020denoising,
	title={Denoising and Regularization via Exploiting the Structural Bias of Convolutional Generators}, 
	author={Reinhard Heckel and Mahdi Soltanolkotabi},
	year={2020},
	eprint={1910.14634},
	archivePrefix={arXiv},
}

@article{curatolo2022recent,
	title={Recent advances in parallel imaging for MRI: WAVE-CAIPI technique},
	author={Curatolo, Calogero},
	journal={Journal of Advanced Health Care},
	volume={4},
	number={1},
	year={2022}
}

@article{tavakkoli2025review,
	title={A Review on Accelerated Magnetic Resonance Imaging Techniques: Parallel Imaging, Compressed Sensing, and Machine Learning},
	author={Tavakkoli, Mitra and Noseworthy, Michael},
	journal={Critical Reviews™ in Biomedical Engineering},
	year={2025},
	publisher={Begel House Inc.}
}

@article{kits20242,
	title={2.5-Minute Fast Brain MRI with Multiple Contrasts in Acute Ischemic Stroke},
	author={Kits, Annika and Al-Saadi, Jonathan and De Luca, Francesca and Janzon, Fredrik and Mazya, Michael V and Lundberg, Johan and Sprenger, Tim and Skare, Stefan and Delgado, Anna Falk},
	journal={Neuroradiology},
	volume={66},
	number={5},
	pages={737--747},
	year={2024},
	publisher={Springer}
}

@article{guo2022emerging,
	title={Emerging techniques in cardiac magnetic resonance imaging},
	author={Guo, Rui and Weing{\"a}rtner, Sebastian and {\v{S}}iuryt{\.e}, Paulina and T Stoeck, Christian and F{\"u}etterer, Maximilian and E Campbell-Washburn, Adrienne and Suinesiaputra, Avan and Jerosch-Herold, Michael and Nezafat, Reza},
	journal={Journal of Magnetic Resonance Imaging},
	volume={55},
	number={4},
	pages={1043--1059},
	year={2022},
	publisher={Wiley Online Library}
}

@article{zhu2023characteristic,
	title={Characteristic-constrained accelerating MR T1rho mapping with blockwise infimal convolution of matrix elastic-net regularization},
	author={Zhu, Qingyong and Cui, Zhuo-Xu and Liu, Yuanyuan and Cheng, Jing and Zhao, Kankan and Wang, Haifeng and Zhu, Yanjie and Liang, Dong},
	journal={Medical Physics},
	volume={50},
	number={4},
	pages={2224--2238},
	year={2023},
}

@article{2016Infimal,
	title={Infimal convolution of total generalized variation functionals for dynamic MRI},
	author={ Schloegl, M.  and  Holler, M.  and  Schwarzl, A.  and  Bredies, K.  and  Stollberger, R. },
	journal={Magnetic Resonance in Medicine},
	volume={78},
	number={1},
	pages={142-155},	
	year={2016},
}

@article{xu2025self,
	title={Self-supervised Deep Unrolled Model with Implicit Neural Representation Regularization for Accelerating MRI Reconstruction},
	author={Xu, Jingran and Liu, Yuanyuan and Yang, Yuanbiao and Cui, Zhuo-Xu and Cheng, Jing and Zhu, Qingyong and Zhang, Nannan and Zhou, Yihang and Liang, Dong and Zhu, Yanjie},
	journal={arXiv preprint arXiv:2510.06611},
	year={2025}
}

@article{zhu2019,
	title={Incorporating reference guided priors into calibrationless parallel imaging reconstruction},
	author={Zhu, Q. Y and Wang, Wei and Cheng, Jing and Peng, Xi},
	journal={Magnetic Resonance Imaging},
	volume={57},
	pages={347-358},
	year={2019},
}

@article{peng2015incorporating,
	title={Incorporating reference in parallel imaging and compressed sensing},
	author={Peng, Xi and Ying, Leslie and Liu, Q. G and Zhu, Y. J and Liu, Y. Y and Qu, X. B and Liu, Xin and
	Zheng, H. R and Liang, Dong},
	journal={Magnetic Resonance in Medicine},
	volume={73},
	number={4},
	pages={1490-1504},
	year={2015},
}

\vfill

\end{document}